\documentclass[11pt]{article}
\usepackage{graphicx}
\usepackage{latexsym,graphicx}
\usepackage{float}
\usepackage{amsmath}
\usepackage{amsfonts}
\usepackage{amssymb}
\usepackage{epstopdf}
\DeclareGraphicsExtensions{.eps}
\usepackage[left=2.5cm,top=2.5cm,right=2.5cm,bottom=2.5cm]{geometry}

\catcode `\@=11
\catcode `\@=12
\begin{document}
	\begin{center}
		\large{\bf{An f(R,T) Gravity Based FLRW Model and Observational Constraints}} \\
		\vspace{5mm}
		\normalsize{Anirudh Pradhan$^1$, Gopikant Goswami$^2$, Rita Rani$^3$, Aroonkumar  Beesham$^{4,5}$}\\
		\vspace{5mm}
		\normalsize{$^{1}$Centre for Cosmology, Astrophysics and Space Science (CCASS), GLA University, Mathura-281 406, Uttar Pradesh, India}\\
        \vspace{5mm}
		\normalsize{$^{2,3}$Department of Mathematics, Netaji Subhas University of Technology, Delhi, India}\\
		\vspace{5mm}
		\normalsize{$^{4}$Department of Mathematical Sciences, University of Zululand Private Bag X1001 
             Kwa-Dlangezwa 3886 South Africa}\\
		\vspace{5mm}
		\normalsize{$^{5}$Faculty of Natural Sciences, Mangosuthu University of Technology, P O Box 
             12363, Jacobs 4052, South Africa}\\
		\vspace{2mm}
		$^1$E-mail: pradhan.anirudh@gmail.com \\
		\vspace{2mm}
		$^2$E-mail: gk.goswami9@gmail.com \\
		\vspace{2mm}
	    $^3$E-mail: rita.ma19@nsut.ac.in \\
		\vspace{2mm}
		$^{4,5}$E-mail: abeesham@yahoo.com \\
	\end{center}

\begin{abstract}

 We attempt to construct a Friedmann-Lemaitre-Robertson-Walker(FLRW) cosmological model in $f(R, T)$ gravity which exhibits a phase transition from deceleration  to acceleration at present.  We take  $f(R,T) = R + 2 \lambda T$, $\lambda$ being an arbitrary constant. In  our model, the $\lambda$ parameter develops a negative pressure in the universe whose Equation of state is parameterized. The present values of model parameters such as density, Hubble, deceleration, Equation of state, and $\lambda$ are estimated statistically by using the Chi-Square test. For this, we have used three different types of observational data sets: the $46$ Hubble parameter data set, the SNeIa $715$ data sets of distance modulus, and the 66 Pantheon data set (the latest compilation of SNeIa 40 bined plus 26 high red shift apparent magnitude $m_b$ data set in the red shift ranges from $0.014 \leq z \leq 2.26 $). We have calculated the transitional red shift and time. The estimated results for the present values of various model parameters are found as per expectations and surveys. Interestingly, we get the present value of the density $\rho_0$, $\simeq 1.5 \rho_c $. The critical density is estimated as $\rho_c\simeq 1.88 ~ h_0^2~10^{-29}~gm/cm^3 $ in the literature. The higher value of the present density is attributed to the presence of some additional energies in the universe apart from baryon energy. We have examined the behavior  of the pressure in our model. It is negative and produces acceleration in the universe. Its present value is obtained as $p_0 \simeq - 0.7 \rho_0$.
\end{abstract}

Observational constraints 

\section{ Introduction}
  $\Lambda$CDM cosmological model is the traditional concordance model which fits best with the latest observational constraints despite its failure to explain fine-tuning and the cosmic coincidence problems (\cite{a} $-$ \cite{d}). A set of observational results are given in  references (\cite{1} $-$ \cite{25}) which expresses the fact that our universe is accelerating. To explain this, a large amount of anti-gravitational and repulsive energy given the name ``exotic dark energy (DE)'' is believed to be present in the universe and this DE is responsible for the acceleration. In observational cosmology surveys, there are searches for mainly four parameters: Hubble parameter ($H_0$), distance modulus ($\mu$), apparent magnitude ($m_b$), and the deceleration parameters (DP - $q_0$). So, these parameters are important tools to model a physical universe. The traditional Friedmann-Lemaitre-Robertson-Walker (FLRW) model is  so far the best-fit model which describes a homogeneous and isotropic universe. It originates with a big bang singularity, then sudden inflation cools down its heavy contents  to permit the production of sub-atomic and quantum particles. Thereafter the universe enters into the radiation and matter-dominated eras. But it fails to explain the higher value of the  density of the universe. It also does not explain why the SNIa supernovae are more distant than expected, which requires  an acceleration in the universe instead of deceleration as predicted by the FLRW model.

  There are two schools of thought to explain and analyze these anomalies. In the former one (\cite{e} $-$\cite{i}), it is assumed that along with baryon  matter, DE  exists producing negative pressure. As a result, it  repels matter,  thus producing acceleration in the universe. DE is discussed in the framework of general relativity. The second school of thought is based on the theme that nonlinear curvature may develop a geometry that could change the dynamics of matter to produce an acceleration in the universe. This requires modifications in Einstein's field equations.  A group headed by A A Stravinsky, Antonio De Felice, and Tsujikawa et al. (\cite{j}$-$ \cite{x})  modified Einstein field equations by replacing the Ricci scalar $ R $ with an arbitrary function of the Ricci scalar $ R $ and the energy-momentum tensor $T_{ij}$ in the Einstein Hilbert action, and formulated modified theories of gravitation. Their views are simple in the sense that matter creates gravitation and gravitation creates curvature. Curvature will not remain silent, it should also act on matter to produce some dynamic results. Accordingly, so many modified theories of gravity $ f(R) $, $ f(R, G) $, $ f(R, T) $ gravity, $ f(R, T^{\phi}) $  and many more have surfaced in the literature. Out of this $ f (R, T), $ is one of the popular options.
 
  In the present work, we attempt to model a universe with reference to the present context in the framework of an FLRW space-time metric using the field equations of $ f(R, T) $ gravity. The propagator of the theory has suggested three options for the specific functional form of $f(R, T)$. We consider the first popular one $f(R, T) = R + 2f(T)$, where we have taken $f(T)= \lambda T$ and $\lambda$ is an arbitrary parameter. The aim is to develop an accelerating model. For this, it is proposed that the $\lambda$ parameter is associated with negative pressure, and the equation of state(EoS) ($\omega$) is parameterized as per Gong and Zhang( \cite{u}). Like the Einstein field equations for an FLRW space-time, we do have a set of two differential equations in which the first one determines acceleration whereas the other one describes the  rate of expansion (Hubble parameter) which involves the density of matter. We have statistically estimated the present values of model parameters, EoS ($\omega_0$), the Hubble ($H_0$), decelerating parameters ($q_0$), and $\lambda$. For this, we consider three types of observational data sets: the $46$ Hubble parameter data set, the SNe Ia $715$ data sets of distance modulus and apparent magnitude, and the 66 Pantheon data set (the latest compilation of SN Ia 40 bined plus 26 high red shift apparent magnitude $m_b$ data set in the red shift ranges from $0.014 \leq z \leq 2.26 $). These sets of data are compared with the theoretical results through the $ \chi^2 $ statistical test and estimated values are obtained on the basis of minimum $\chi^2$. The model exhibits a phase transition from deceleration  to acceleration. We have calculated transitional red shifts and time for the data sets. Our estimated results for the present values of various model parameters such as the Hubble, deceleration, etc., are found as per expectations and surveys. The higher value of the present density is attributed to the  presence of  additional energies in the universe apart from baryon energy. We have also examined the behavior  of the pressure in our model. It is negative and produces an acceleration in the universe. Its present value is obtained as $p_0 \simeq - 0.7 \rho_0$.\\
    
 The outline of the paper is as follows: In section II, the $ f(R, T) $ gravity field equations along with the action and the three specific functional forms of $f(R, T) $ are described. In sec. III, the $ f(R, T) $ field equations are obtained for the linear form of $ f(R, T) = R+2 \lambda T $ in the framework of the FLRW spatially flat space-time. In this section, we have solved the field equations to find the expressions for the Hubble and deceleration parameters. In section IV, the distance modulus, luminosity distance, and apparent magnitude are defined and formulated. Statistical estimation and evaluation of the model parameters are done in sections V and VI. In these sections, we have plotted various error bars, and likelihood graphs, the 1$\sigma$ and 2$\sigma$ confidence regions and  the deceleration parameter $ (q) $, jerk parameter $ (j) $ and snap parameter $ (s) $ versus red shift ($z$) graphs. We have obtained transitional red shifts and corresponding times which display how the universe passed from the deceleration to the acceleration era. In section VII, a state finder analysis is carried out which tells us that our model at present is in quintessence and its evolution passed through the Einstein - de Sitter and $ \Lambda $CDM stages. In the last section, we have summarized the work with the conclusion.
\section{f(R,T) gravity} 

The Einstein field equations (EFE) are given by:

\begin{equation}\label{1}
R_{ij}-\frac{1}{2} R g_{ij}+ \Lambda g_{ij} = \frac{8\pi G}{c^4}T_{ij},
\end{equation}
where the symbols have their usual meanings.
These Eqns. are obtained from the following action:
\begin{equation}{\label{2}}
	S= \int(\frac{1}{16\pi G}( R+2\lambda) + L_m)\sqrt{-g} dx^4.
\end{equation}
Harko et al. \cite{t}  modified the GRT field equations by replacing the Ricci scalar R with an arbitrary function $ f(R, T) $ of $R$ and the trace $T$ of the energy-momentum tensor $T_{ij}$.
The action for $f(R,T)$ gravity is:
\begin{equation}{\label{3}} 
 S= \int\Big(\frac{1}{16\pi G} f(R,T)+L_m\Big)\sqrt{-g} dx^4,
\end{equation}
where  $ L_m $ denotes the matter Lagrangian density. 
The stress-energy tensor of the matter is defined as \cite{s}:
\begin{equation}{\label{4}} 
    T_{ij}= -\frac{2}{\sqrt{-g}}\frac{\delta(\sqrt{-g} L_m)}{\delta g^{ij}}
\end{equation}
By taking the variation of the action $ S $ with respect to the metric tensor components $ g_{ij} $, the field equations of $ f(R, T) $ gravity are obtained as \cite{t}:
\begin{equation} {\label{5}}
    R_{ij}-\frac{1}{2} R g_{ij} = \frac{8 \pi G T_{ij}}{f^R(R,T)}+ \frac{1}{f^R (R,T)} \bigg(\frac{1}{2} g_{ij} (f(R,T)-R f^R (R,T)) - (g_{ij} \Box - \nabla_i \nabla_j) f^R (R,T) + f^T (R,T) (T_{ij} +p g_{ij}) \bigg),
\end{equation}
where $ f^R $ and $ f^T $ denote the derivatives of $ f(R,T) $ with respect to $ R $ and $ T $, respectively. The Lagrangian for a perfect fluid is $ L_m = - p $, and its energy momentum tensor is:
\begin{equation}\label{6}
 T_{ij}= (\rho + p) u_i u_j - p g_{ij},
\end{equation}
where $ \rho $ and $ p $ are the energy density and pressure, respectively. The vector $ u_i = (0,0,0,1) $ is the four-velocity in the co-moving coordinate system which satisfies the conditions $u^{i} u_{i} =1$ and $u^{i} u_{i;j} =0$. In \cite{t}, the authors proposed the following three cases for the function $f(R, T)$  for cosmological applications:
\begin{displaymath}
	f(R,T) = \left\{
	\begin{array}{lr}
		R + 2f(T)\\
		f_{1}(R) + f_{2}(T)\\
		f_{1}(R) + f_{2}(R)f_{3}(T)
	\end{array}
	\right\}.
\end{displaymath}
Numerous authors \cite{t1,t2,t3,t4,t5} have recently examined in detail the cosmological implications for the class $f(R, T) = R + 2f(T)$.
According to Fisher and Carlson's recent study of $f(R, T)$ gravity \cite{t6}, the term $f_{2}(T)$ should be included in the matter Lagrangian $L_{m}$, and hence has no physical meaning. They concentrated especially on the scenario where $f$ is separable, resulting in $f(R,T)= f_{1}(R) + f_{2}(T)$. Harko and Moraes \cite{t7} thoroughly reexamined the findings of the paper \cite{t8} and demonstrated that their physical analyses and interpretation of the $T$-dependence of $f(R, T)$ gravity contained significant conceptual problems. We refer to recent publications \cite{t9,t10,t11,t12,t13,t15,t16} for a better grasp of the cosmological implications and mathematical structure of $f(R, T)$ gravity. We plan to investigate a cosmological model  based on $f(R, T)$ theory which fits  best with current observations, and which can be compared with the findings of the  $\Lambda$CDM model. So we take the simple linear case $f(R, T)= R + \lambda T $ where $\lambda$ is a scalar that couples $R$ and $T$.

\section{Metric and Field Equations } 
The FLRW spatially flat space-time is given as:
 \begin{equation}{\label{7}}
 ds^{2} = dt^{2} - a^{2}(t) (dx^{2} + dy^{2} + dz^{2}),
 \end{equation}
where $ a(t) $ represents the scale factor. The trace of the stress energy-momentum tensor is obtained as:
\begin{equation}{\label{8}}
 T= \rho - 3 p
\end{equation}
The $ f(R,T) $ field equations (\ref{5}) for the metric (\ref{7}) are obtained as:
\begin{equation}{\label{9}}
  2 \dot{H} + 3 H^2 = -(8 \pi + 3 \lambda) p + \lambda \rho
\end{equation}
and
\begin{equation}{\label{10}}
  3 H^2 = (8 \pi + 3 \lambda ) \rho - \lambda p,
\end{equation}
where the Hubble parameter, $H=\frac{\dot{a}}{a}$.
We  assume $ \lambda = 8 \pi \eta $. Then the field equations (\ref{9}) and (\ref{10}) are written as:
\begin{equation}{\label{11}}
  (1 - 2 q) H^2 = 8 \pi ( \eta \rho-(1 + 3 \eta) p )
\end{equation}
and
\begin{equation}{\label{12}}
  3 H^2 = 8 \pi  ( ( 1 + 3 \eta ) \rho - \eta p ),
\end{equation}

 where $q=-\frac{\ddot{a}}{a H^2}$ is deceleration parameter.
\\

As the universe is currently accelerating, so both the  deceleration parameter and pressure must be negative. The observations tell us that the luminous content of the universe (baryon fluid) is dust at present so the baryon pressure must be zero. But the literature \cite{e} says  that apart from baryon matter, other forms of matter do exist in the universe. It is estimated that nearly $ 28 \% $  and  $ 68 \% $ of the total content of the universe is the dark matter and dark energy  respectively. The dark matter is responsible for the phenomenon of gravitational lensing and dark energy is for the present-day acceleration in the universe. These ideas and how to accommodate them in  theories have been explained in the introduction. In $ f(R, T) $ gravity, the Ricci scalar $ R $ is replaced by an arbitrary function of  $ R $ and  $T$. The idea behind is that to have acceleration due to curvature and trace dominance. The authors feel that the pressure term arising in the field equations is not due to the baryon matter but a result of the overall effect. We mean that terms containing $ \eta $ in the field equations (\ref{11}) and (\ref{12})  are the extra terms in the original FLRW field equations of general relativity, and they will have an impact in producing pressure and creating acceleration in the universe. We observe that there are two Eqs. (\ref{11}) and (\ref{12})  with four unknowns, viz., $H$, $q$, $p$ and $\rho$. Therefore, to get an explicit solution to the above equations, we need to assume at least one reasonable relation among the variables or we may parameterize the variables. For this, we assume the usual equation of state for the fluid as $ p  = \omega~ \rho $, and we consider the parameterization of the equation of state parameter ($\omega $) as given by Gong and Zhang\cite{u}:
  $$\omega = \frac{\omega_0}{(1+z)}, $$ where $z$ is the red shift and $\omega_0$ is the present value of $\omega$.
 
 Eq. (\ref{12}) is re-written in the following form:
  \begin{equation}\label{13}
 \frac{8 \pi\rho_0}{3 H_0^2}=\frac{1}{1+k}, k= \eta(3-\omega_0), \omega_0= 3- \frac{k}{\eta}
 \end{equation} \\
 
 where the suffix $0$ denotes the present values of the parameters.
 From this, we can find $$\rho_0 = \frac{3H^2_0}{8\pi(1+k)}=\frac{\rho_c}{1+k},$$ where $\rho_c =\frac{3H^2_0}{8\pi} $ is critical density.
Eqs. (\ref{11}) and (\ref{12} ) are simplified  as follows:
\begin{equation}{\label{14}}
  (1 - 2 q) H^2 = 8 \pi \rho \left( \eta -\frac{ (1 + 3 \eta) (3 - \frac{k}{ \eta})} {1+z}\right) 
\end{equation}
and
\begin{equation}{\label{15}}
  3 H^2 = 8 \pi \rho \left( ( 1 + 3 \eta )  -\frac{\eta (3 - \frac{k}{ \eta})} {1+z}\right ).
\end{equation}
From Eqs. (\ref{14}) and (\ref{15}), we get 
\begin{equation}{\label{16}}
\frac{1-2 q_0}{3}  = \frac{ -(1+3 \eta) (3 - \frac{k}{ \eta}) + \eta}{(1+3 \eta) - (3 - \frac{k}{\eta}) \eta}
\end{equation}
We solve this equation for $ \eta $  in terms of $ q_0 $ and $ k $, we  get
 \begin{equation}{\label{17}}
     \eta  = \frac{1}{48} \left(2 k q_0 + \sqrt{(-2 k  q_0 - 8 k-2  q_0  +10)^2+288 k}+8 k+2  q_0 -10\right)
\end{equation}
\section{Derivations of various cosmological parameters: }
\subsection{Deceleration, Jerk, and Snap Parameters:}

From  Eqs. (\ref{14}), (\ref{15}) \&  (\ref{17}), the  deceleration parameter $q$ is solved as a function of red shift $z$ in terms of the parameters $q_0$ and $k$ as follows:
\begin{equation}{\label{18}}
		q(z) = \frac{8 (k+1) q_0+4 z}{z (k (q_0+4)+ q_0+3)+z \sqrt{(k ( q_0 + 4)+ q_0 - 5)^2 + 72 k} + 8 (k+1)}.
		\end{equation}
  
There are two more parameters, the jerk ($j$), and snap($s$) which are related to the  third and fourth-order derivatives of the scale factor. They play a very important role in examining the instability of a cosmological model. They are defined as follows, $j=\frac{\dddot{a}}{a H^3}$ and  $s= -\frac{\ddddot{a}}{a H^4}. $
The jerk parameter in the terms of deceleration parameter can be written as: 
        \begin{equation}{\label{19}}
		j(z) = q(z) +2 {q(z)}^2 + (1+z) \frac{d q(z)}{d z}.
	\end{equation}
Then using Eq. (\ref{18}) in Eq. (\ref{19}), the jerk parameter in terms of $ k $ and $ q_0 $ have following forms in our model:
\begin{displaymath}
    j(z) = \left\{
		\begin{array}{lr}
			(A+B)/\left(z (k (q_{0}+4)+ q_{0}+3)+z \sqrt{(k (q_{0}+4)+q_{0}-5)^2+72 k}+8 (k+1)\right)^2\\A=4 z^2 (k (q_{0}+4)+q_{0}+11)+64 (k+1) (2 {q_0}+1) z+8 (k+1) (q_{0} (15 (k+1)q_{0}+4 k+5)+4)\\B=4 \sqrt{(k (q_{0}+4)+q_{0}-5)^2+72 k} \left(z^2-2 (k+1) {q_0}\right)
		\end{array}
		\right\}.
	\end{displaymath}
     \begin{equation}\label{20}
     \end{equation}
The snap parameter in terms of the deceleration and jerk parameters is computed as:
\begin{equation}{\label{21}}
		s(z) = ( 3 q(z) +2 ) j(z) + \frac{d j(z)}{d z} (1+z)
\end{equation}


\subsection{Hubble Parameter:}
The Hubble parameter is related to the deceleration parameter $q$ in the equation given below,

\begin{equation}\label{22}
H_z(1+z)=(q+1)H	
\end{equation}

Using Eq. (\ref{18}) and integrating Eq. (\ref{22}), we get the following expression for the Hubble parameter: 

\begin{equation*}
	H = H_0 e^{\int_{0}^{z}\frac{(q+1)dz}{(1+z)}}
\end{equation*}
and
\begin{displaymath}
H = \left\{
\begin{array}{lr}
	H_0 (z+1)^m
	\left(1+\frac{\left( \sqrt{k^2 (q_{0}+4)^2+2 k ((q_{0}-1) q_{0}+16)+(q_{0}-5)^2}+q_{0}+3+k (q_{0}+4)\right)z}{8(1+k)}\right)^l\\l=\frac{4 (k+1) \left(\sqrt{k^2 (q_{0}+4)^2+2 k ((q_{0}-1) q_{0}+16)+(q_{0}-5)^2}+k (q_{0}+4)-q_{0}+5\right)}{k \left(\sqrt{k^2 (q_{0}+4)^2+2 k ((q_{0}-1) q_{0}+16)+(q_{0}-5)^2}+k (q_{0}+4)+q_{0}+3\right)}\\m=\frac{-\sqrt{k^2 (q_{0}+4)^2+2 k ((q_{0}-1) q_{0}+16)+(q_{0}-5)^2}+k (q_{0}-2)+q_{0}-5}{2 k}\\
\end{array}
\right\}.
\end{displaymath}
\begin{equation}\label{23}
\end{equation}

\subsection{Luminosity Distance, Distance Modulus  and  Apparent Magnitude:}
The luminosity distance-red shift relation is a useful tool to explore the evolution of the universe. As the universe expands, the light originating from a distant luminous body becomes red-shifted. The flux of a source is determined using the luminosity distance. It is described as

\begin{equation}\label{28}
	D_{L}= a_{0} (1+z) r
\end{equation}

The radial coordinate of the source is indicated by $r$ here.
The following shows how the luminosity distance relates to the Hubble parameter:

\begin{equation}\label{29}
	D_{L}= (1+z) a_{0} c\int_{0}^{z}\frac{dz}{H(z)}.
\end{equation}

It is an increasing function of redshift.

There is another  useful observable parameter, the distance modulus $\mu$ which is  related to the luminosity distance  by the following formula:

\begin{equation}\label{30}
	\mu =  m_{b} - M = 5log_{10}\frac{D_{L}}{Mpc}+25,
\end{equation}

where $m_{b}$ and $M$ are the apparent and absolute magnitude of the source, 
respectively. The luminosity distance $D_{L}$ for a supernova at a very small redshift is approximated as

\begin{equation}\label{31}
	D_{L} =\frac{cz}{H_{0}}.
\end{equation}

From  Eqns. (\ref{30}) and (\ref{31}), the absolute magnitude of a low red shift  supernova($z=0.014$) of $m_b=14.57 $ is obtained as:

\begin{equation}\label{32}
	M =  14.57 + 5log_{10}\left(\frac{H_0 Mpc}{0.014 c}\right) - 25.
\end{equation}

From Eqs. (\ref{30})$-$ (\ref{32}), the  absolute magnitude of a supernova is obtained as:

\begin{equation}\label{33}
	m_{b} = 14.57+ 5log_{10}\left[\frac{1+z}{.014c}\int_{0}^{z}\frac{dz}{h(z)}\right],
\end{equation}
where $ h(z)= \frac{H(z)}{H_0}.$


\section{Estimations of model parameters $H_0$, $q_{0}$, $\lambda$ and EOS(Equation of state) $\omega_0$ from various  observational data set:}

\subsection{Estimation by $46$ Hubble data set}
We use a data set of $46$ Hubble parameter observed values for different redshifts in the range ({$0 \le z \le 2.30$}) along with possible standard errors. The data set is displayed in Table 1, in the Appendix. This data set is used to estimate the parameters $H_0$, $q_0$, and $k$ to get correct expressions for $H$, which fits best with the observations. The estimation is done with the help of the following chi-square function:

\begin{equation}\label{34}
\chi^{2}( H_0, k, q_0) = \sum\limits_{i=1}^{46}\frac{[Hth(z_{i},H_0, k, q_0) - H_{ob}(z_{i})]^{2}}{\sigma {(z_{i})}^{2}},
\end{equation}

where, $ H_{ob} $ and $ Hth $ are the observed and theoretical values of $ H $. $ \sigma{(z_{i})} $ denotes the standard error in  $H_{ob}$.
The estimation is done by finding the minimum $\chi^{2}$ over the ranges $({H_0}, 65, 75)$, $(k,-1,0)$ and $(q_0,-1,0 )$ of the model parameters $H_0$, $ k $ and $q_0$.
The estimated values are obtained as  $ H_0 = 67.786_{-1.5142}^{+1.4967} $, $ k = -0.318773_{-0.05321}^{+.04145} $ and $ q_0 = -0.530244_{-0.03451}^{+.03234}  $ for the minimum $ \chi^2 =  21.6233$.
From these we get $\omega_0=-0.742859$ and  $ \lambda = -0.003387$. The current value of the density $\rho_0 = 1.46794 \rho_c $. The critical density is estimated as $\rho_c\simeq 1.88~ h_0^2~10^{-29}~gm/cm^3 $ in the literature.
We present the following  Figures (\ref{Fig-1}) and (\ref{Fig-2}) to show the estimations graphically. These include  1$\sigma$ and 2$\sigma$ confidence regions for the pair of model parameters  ($H_0$,$q_0$), likelihood plots for the parameters $H_0$ and $q_0$ and error bar plots for the Hubble parameter $H$ and rate of expansion $ \frac{H}{1+z}\propto \dot{a} $.


\begin{figure}[H]
	\includegraphics[width=9cm,height=8cm,angle=0]{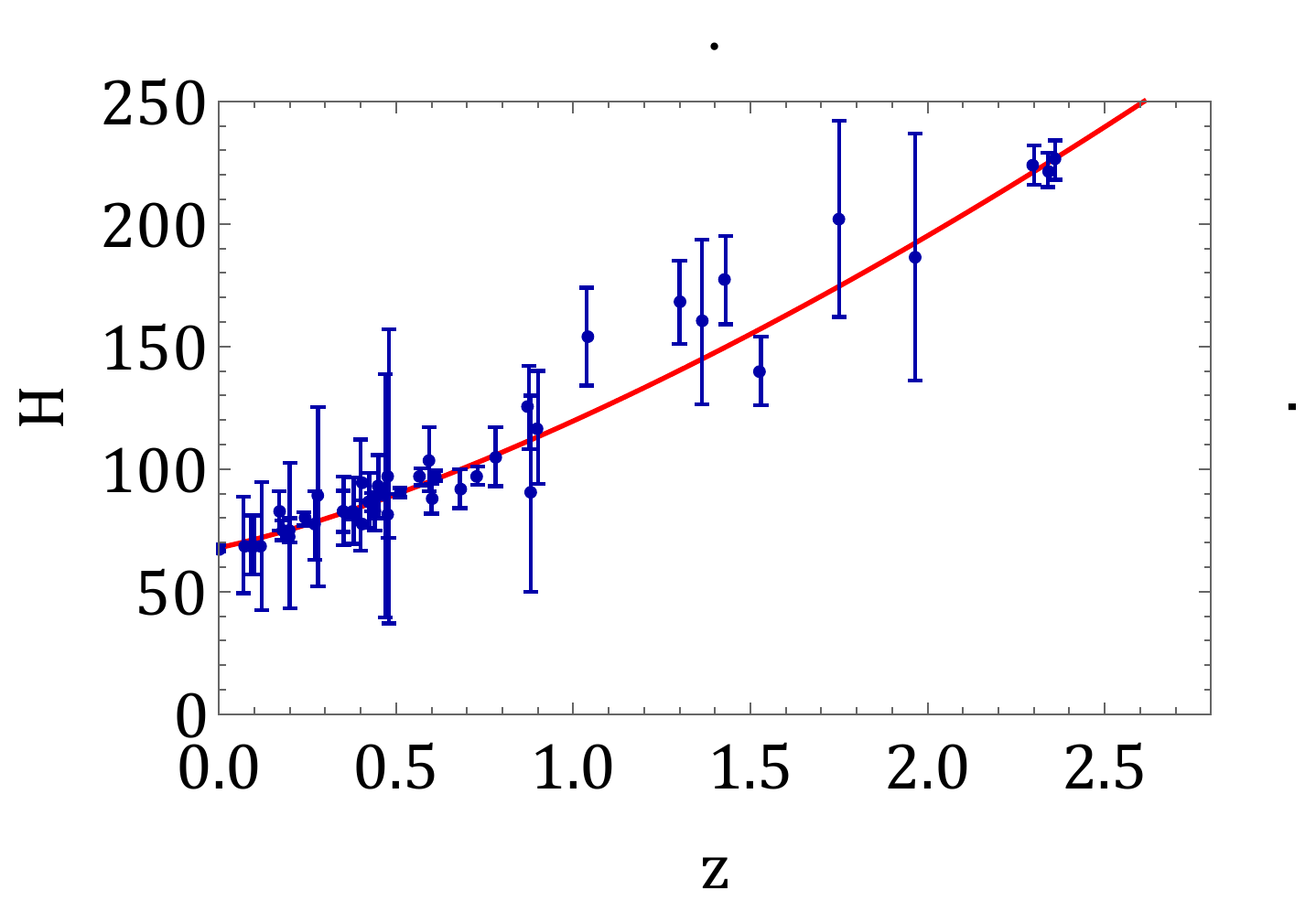}
	\includegraphics[width=9cm,height=7cm,angle=0]{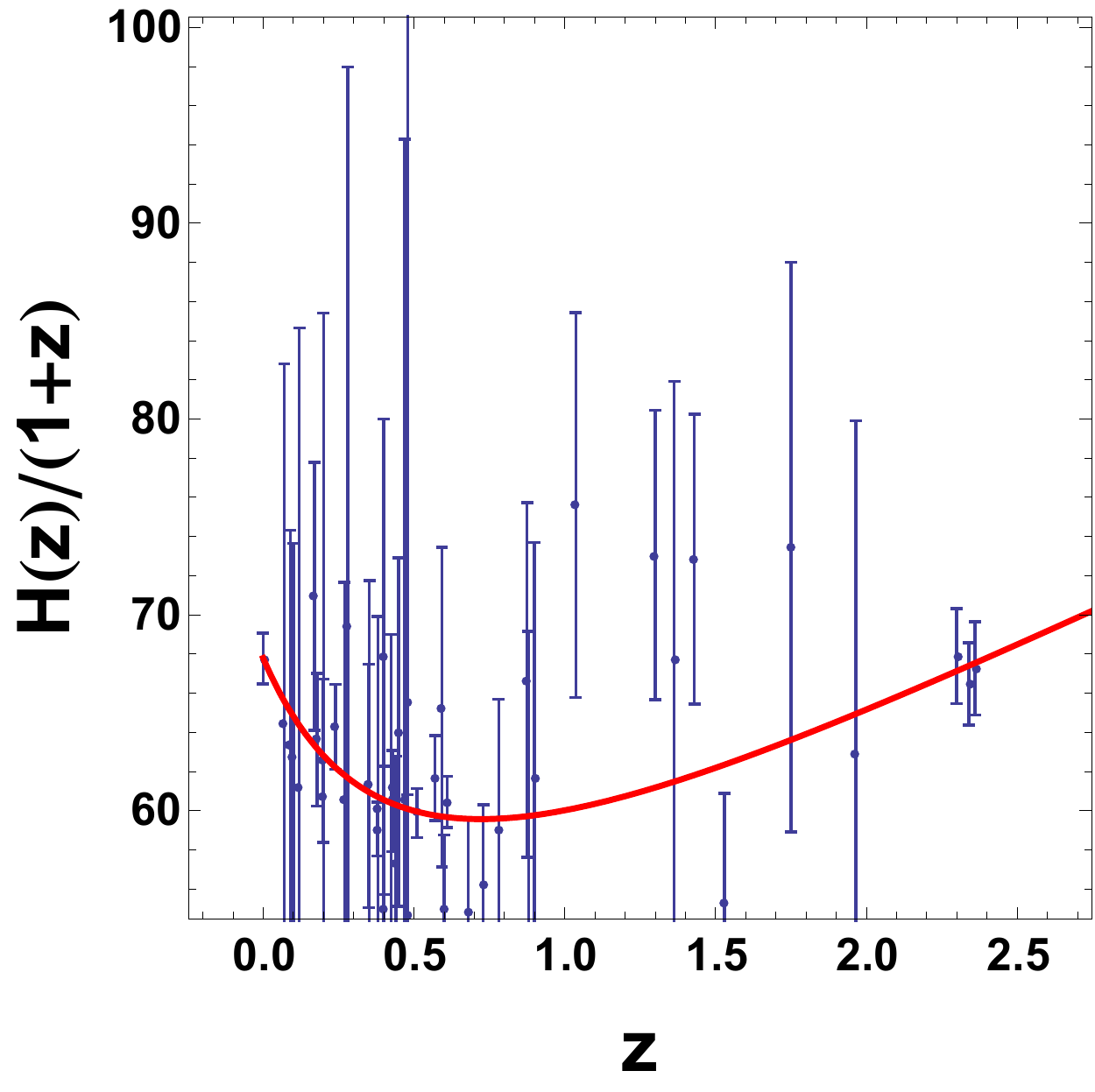}
	\caption{Error bar plots for the Hubble parameter $H$ and the rate of expansion $\dot{a}\propto \frac{H}{1+z}$ using  46 Hubble data set.}
 \label{Fig-1}
\end{figure}


\begin{figure}[H]
(a){\includegraphics[width=9cm,height=8cm,angle=0]{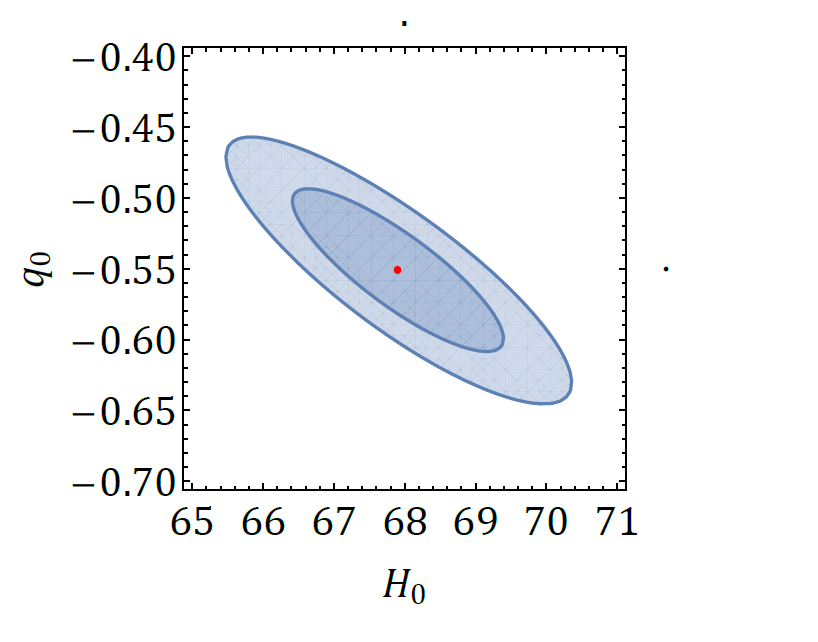}}
(b){\includegraphics[width=9cm,height=8cm,angle=0]{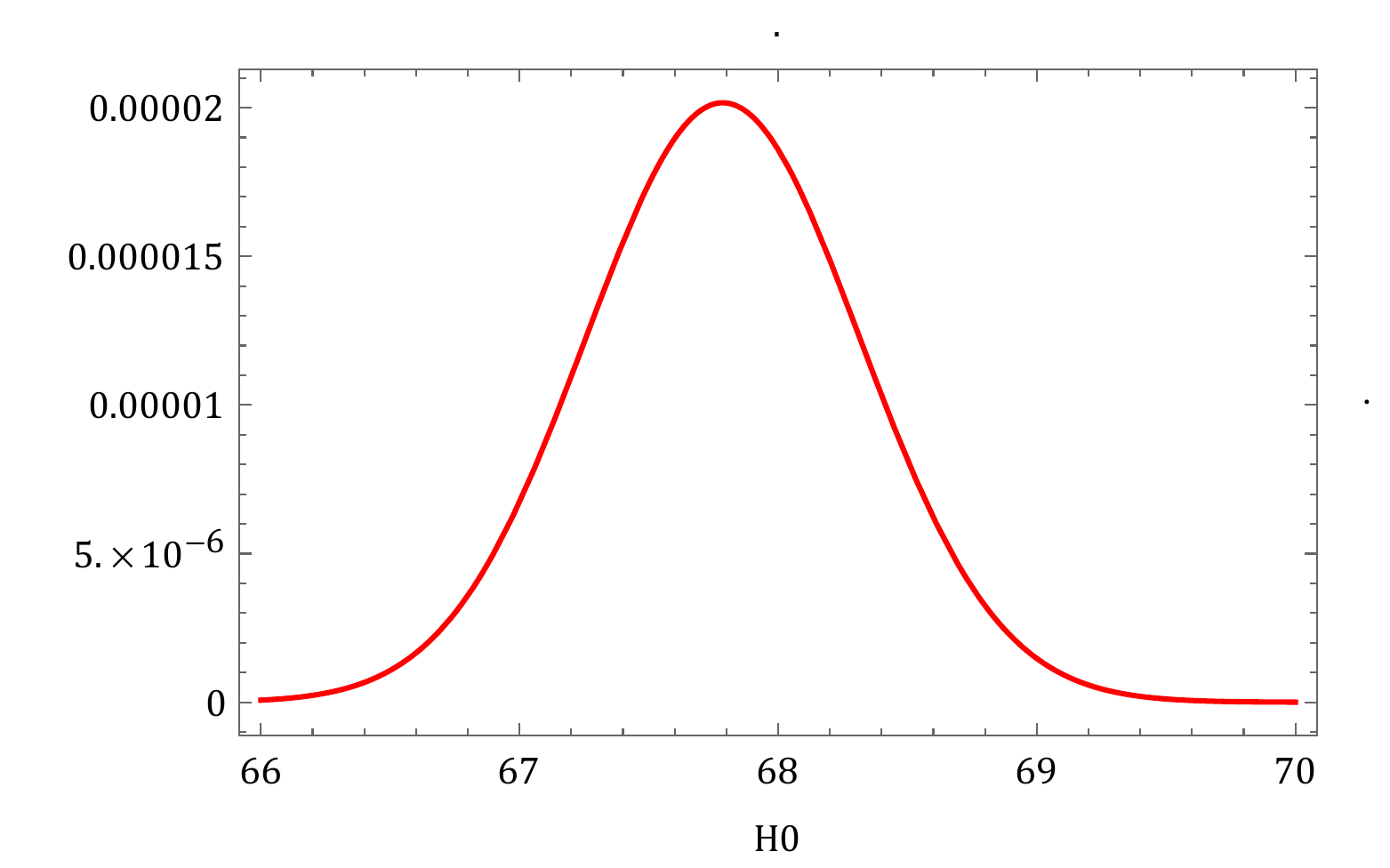}}\\
(c){\includegraphics[width=9cm,height=8cm,angle=0]{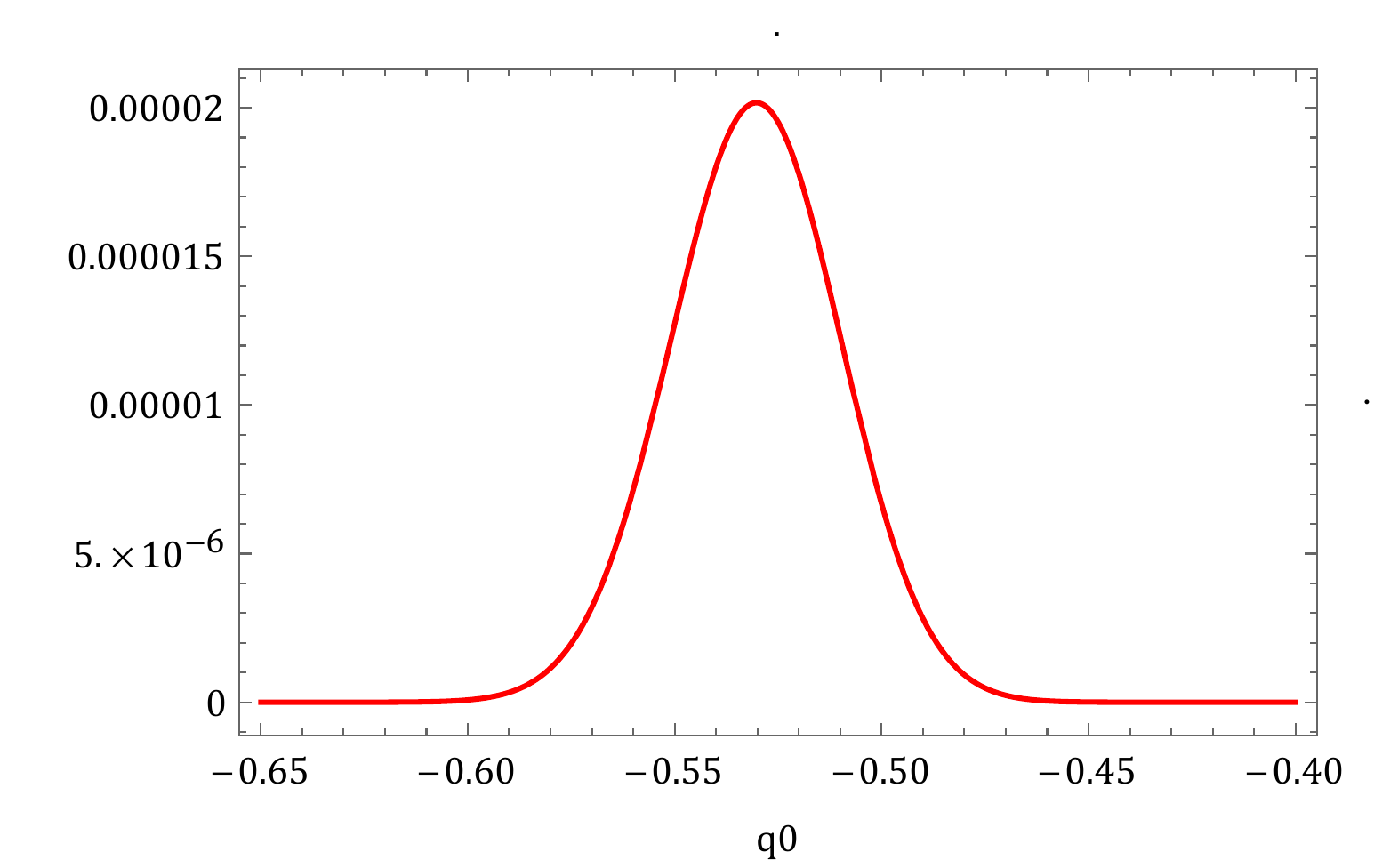}}
\caption{Figure (a) is the contour plot showing 1$\sigma$ and 2$\sigma$ confidence regions for estimated values of model parameters $H_0$ and $q_0$. Figures (b) and (c) are the likely hood curves for $H_0$ and $q_0$. The estimated values $ H_0 = 67.786$ and $ q_0 = -0.530244 $ are at the top. }
\label{Fig-2}
\end{figure}

 \subsection{Estimation by SNeIa $715$ Distance modulus $\mu$ data set}
We solve numerically  Eqs. (\ref{29}) and (\ref{30}) and by using the $715$ SNeIa  distance modulus $\mu$ data set, we estimate the model parameters $H_0$, $q_0$ and $k$ to get a correct expression for $\mu$, which fits best with the observations. The estimation is done with the help of the following chi-square function:

\begin{equation}\label{35}
\chi^{2}(H_{0}, k, q_{0}) = \sum\limits_{i=1}^{715}\frac{[\mu th (z_{i},H_{0}, k, q_{0}) - \mu ob(z_{i})]^{2}}{\sigma {(z_{i})}^{2}},
\end{equation}

where $  \mu ob $ and $\mu th $ are the observed and theoretical values of $\mu $. The quantity  $\sigma{(z_{i})} $ denotes the standard error in  $ \mu ob$.
The estimation is done by finding the minimum $\chi^{2}$ over the ranges $(H_0, 65, 75)$, $(k,-1,0)$ and $(q_0,-1,0 )$ of the model parameters $H_0$, $k$ and $q_{0}$.The estimated values are obtained as $ H_{0}=69.5679 $, $ k = -0.362656 $ and
$ q_{0} = -0.597875 $ for the minimum  $\chi^2 $=  $1220.51$.
Like the previous Hubble data set, we get $\omega_0=-0.78868$ and  $\lambda = -0.00380862$. The present value of density $\rho_0 = 1.56901 \rho_c $. The higher values of the present density attributes to the  presence of the $ \lambda$ terms which create pressure and accelerate the universe.

We present the following  fig (\ref{Fig-3}) to show the estimations graphically. These include the contour plots showing  1$\sigma$ and 2$\sigma$ confidence regions for the estimated values of the model parameters $H_0$ and $q_0$, likelihood plots for $H_0$ and $q_0$, and the error bar plot for the distant modulus $\mu$ using the $715$ SNeIa distance modulus $\mu$ data set.
\begin{figure}[H]
(a){\includegraphics[width=6cm,height=6cm,angle=0]{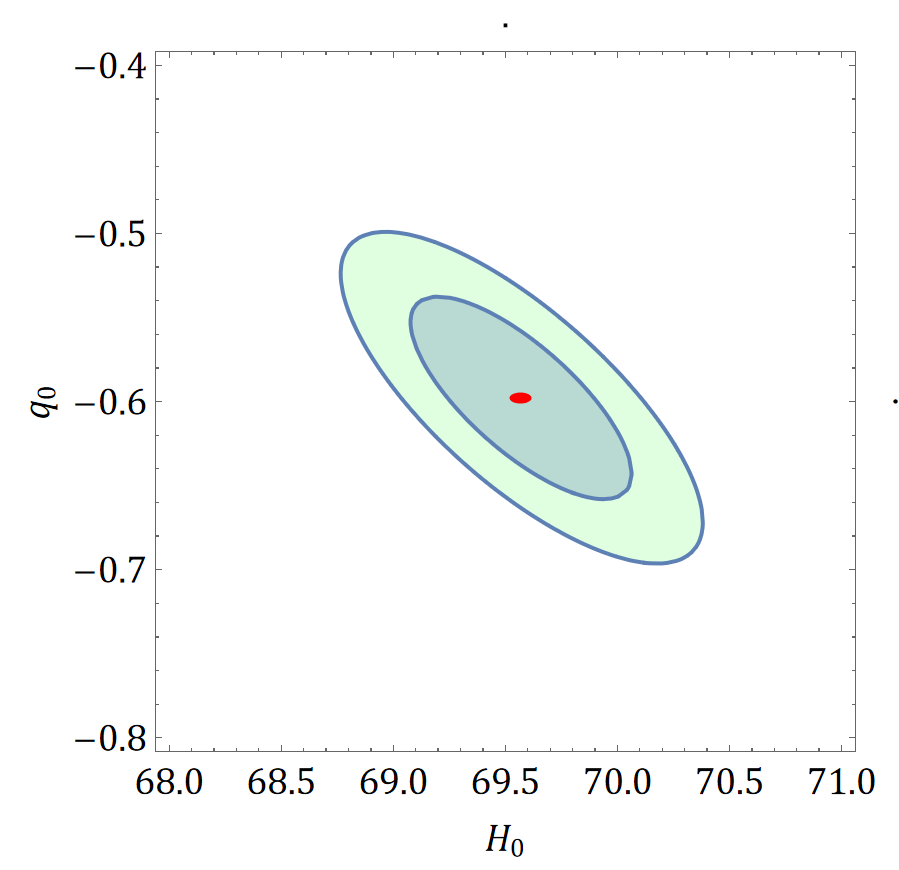}}\hfill
(b){\includegraphics[width=6cm,height=6cm,angle=0]{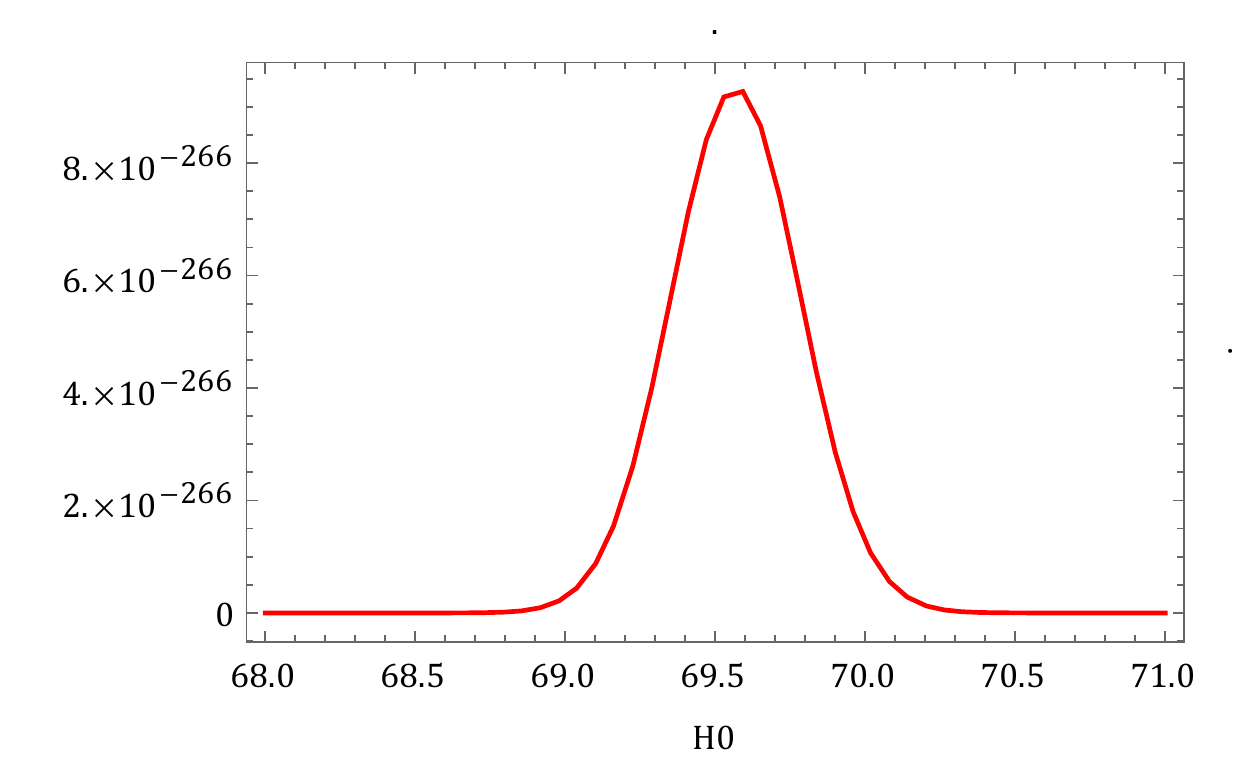}}\par
(c){\includegraphics[width=6cm,height=6cm,angle=0]{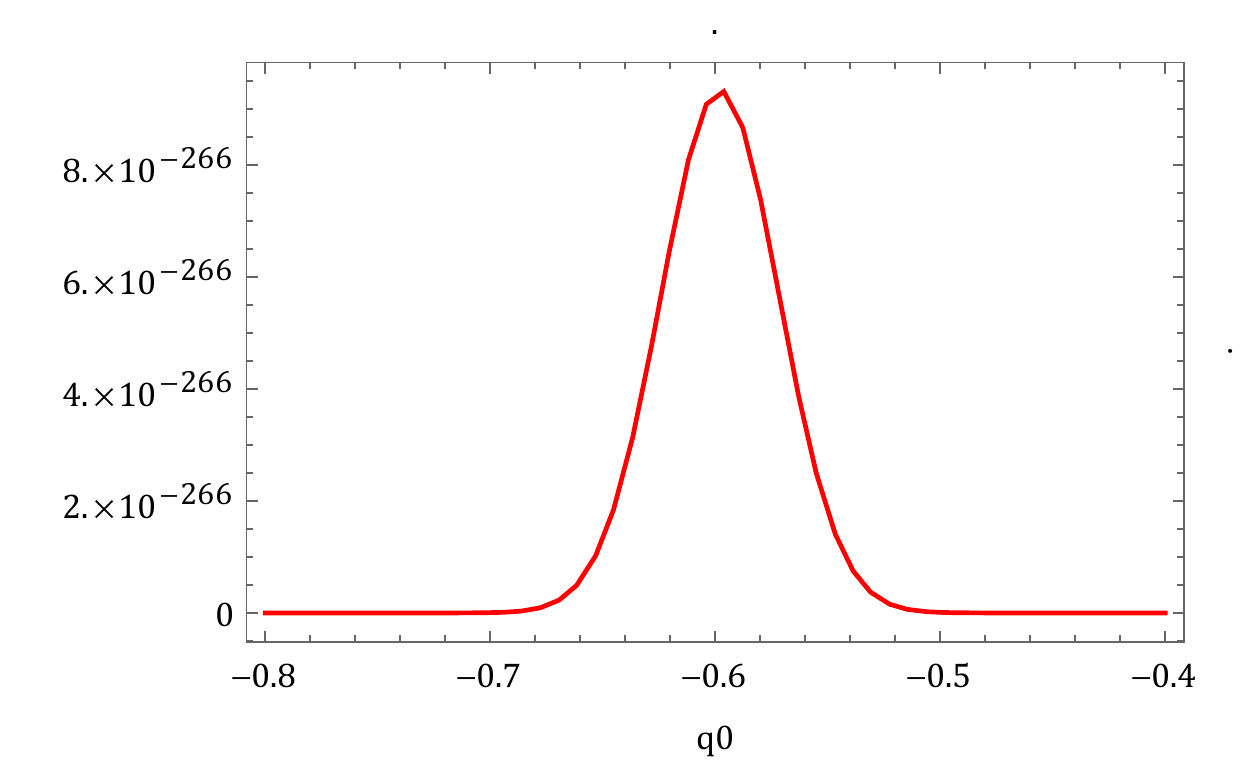}}\hfill
(d){\includegraphics[width=6cm,height=6cm,angle=0]{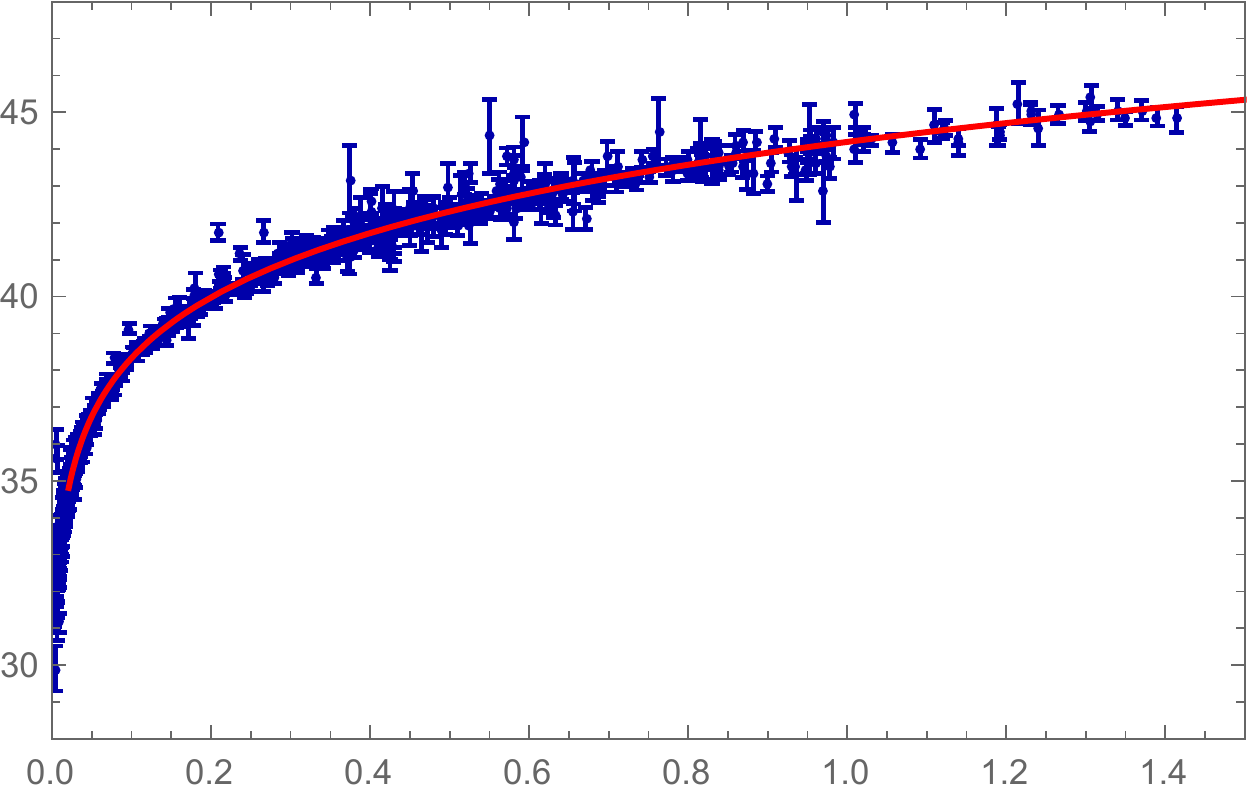}}
\caption{Figure (a) is the contour plot showing  1$\sigma$ and 2$\sigma$ confidence regions for the estimated values of the model parameters $H_0$ and $q_0$. Figures (b) and (c) are the likelihood plots for $H_0$ and $q_0$. The estimated values $ H_0 = 69.5679$ and $ q_0 =  -0.597875 $ are at the top. Figure (d) is the error bar plot for the distant modulus $\mu$ using the $715$ SN Ia distance modulus $\mu$ data set.}
\label{Fig-3}
\end{figure}

\subsection{Estimation by Pantheon $40$ bined plus $26$ high red shift apparent magnitude $m_b$ data sets}
We solve numerically  Eq. (\ref{33}) and by using the $66$
Pantheon data set (the latest compilation of SNeIa $40$ bined plus $26$ high
redshift apparent magnitude $m_b$ data set in the red shift range $0.014 \leq z \leq 2.26 $). This data set may be used to estimate the parameters $q_{0}$ and $k$ to get the expressions for $m_b$ which fits best with the observations. The estimation is done with the help of the following chi-square function:
\begin{equation}\label{36}
\chi^{2}( k, q_0) = \sum\limits_{i=1}^{66}\frac{[m_b th (z_{i}, k, q_0) - m_{b} ob(z_{i})]^{2}}{\sigma {(z_{i})}^{2}},
\end{equation}
where $  m_{b} ob $ and $m_b th $ are the observed and theoretical values of $ m_b $. $ \sigma{(z_{i})} $ denotes the standard error in  $ m_{b} ob$.
The estimation is done by finding the minimum $\chi^{2}$ over the ranges  $(k,-1,1)$ and $(q_{0},-1,0 )$ of the f model parameters $k$ and $q_0$. The estimated values are obtained as $ k = 0.0836673 $ and $ q_{0} = -0.440048 $ for the minimum
$ \chi^2 $=  $95.4103$.

From this, $\omega_0 = -0.61337$ and $ \lambda = 0.001$, and the present value of density $\rho_0 = 0.922792 \rho_c $. We present the following  fig. (\ref{Fig-4}) to show the estimations graphically. These include the contour plot showing  1$\sigma$ and 2$\sigma$ confidence regions for the estimated values of the model parameters $k$ and $q_0$, the likelihood plots for  $k$ and $q_0$ and  the error bar plot for apparent magnitude $m_b$ using the Pantheon 66 (40 bined plus 26 high red shift) SN Ia  $m_b$ data set.


\begin{figure}[H]
a.\includegraphics[width=9cm,height=8cm,angle=0]{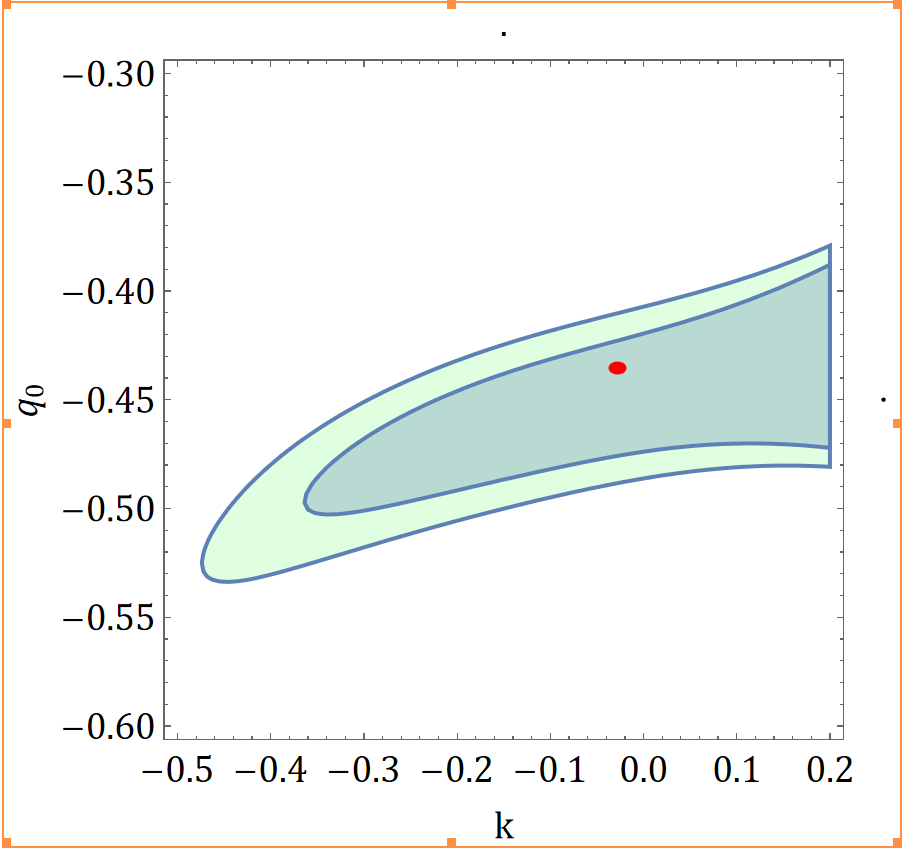}
b.\includegraphics[width=9cm,height=8cm,angle=0]{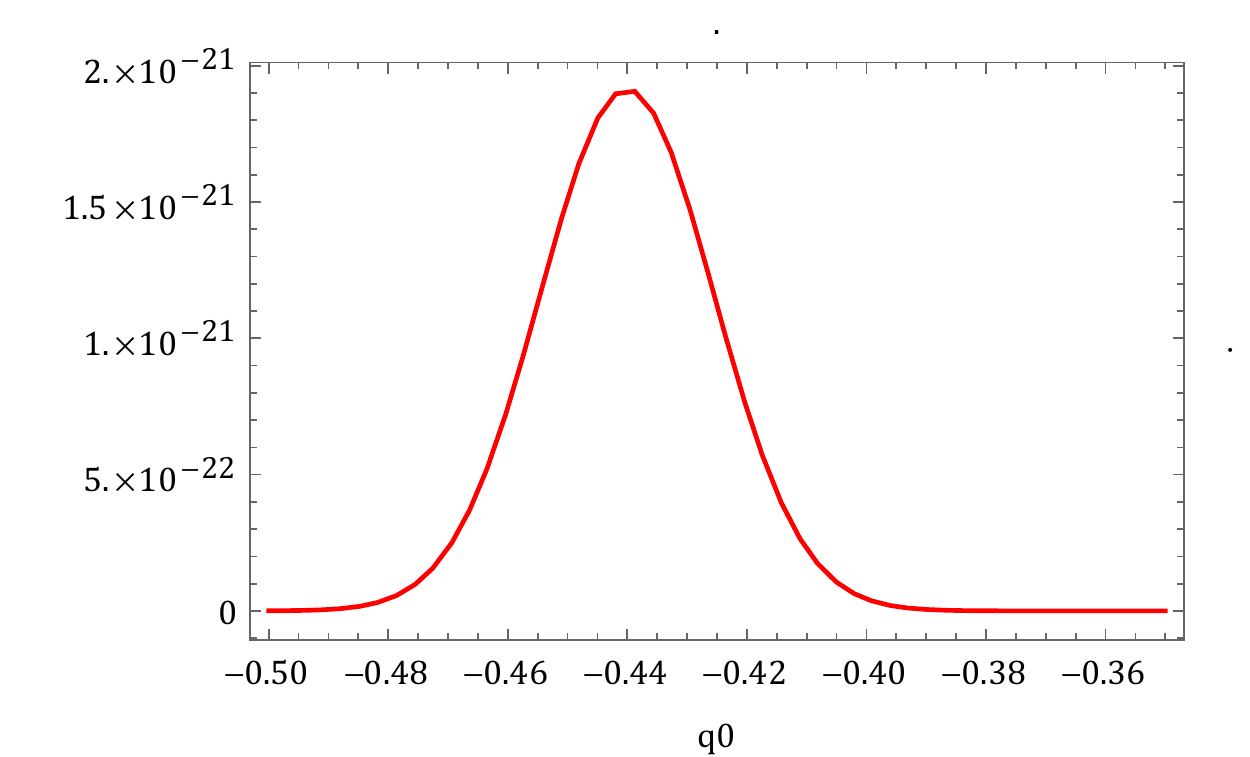}
c.\includegraphics[width=9cm,height=8cm,angle=0]{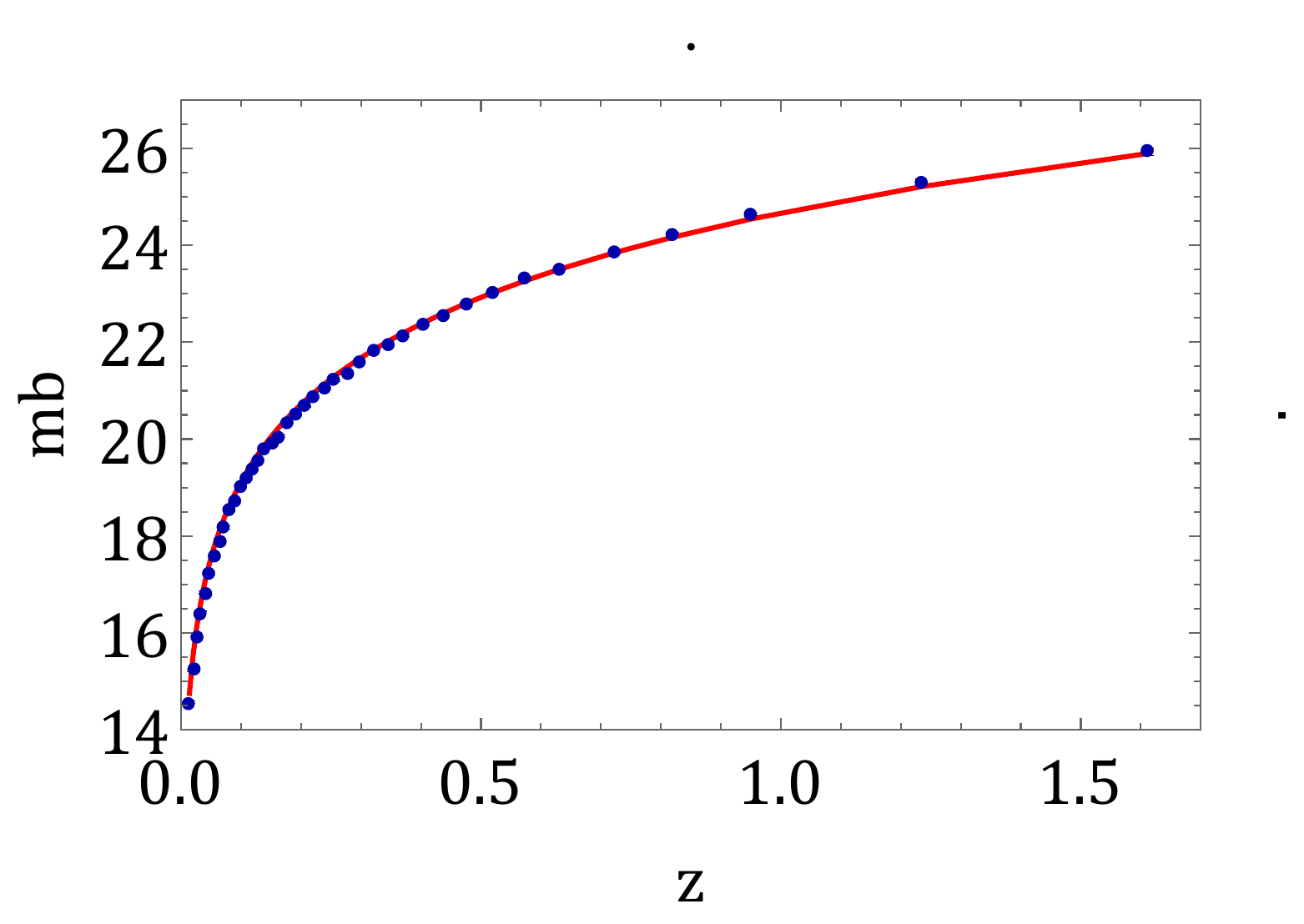}
\caption{Figure (a) is the contour plot showing  1$\sigma$ and 2$\sigma$ confidence regions for the estimated values of the model parameters $k$ and $q_0$. Figures (b) are the likelihood plot for $q_0$. The estimated value  $ q_0 =  -0.440048 $ is at the top. Figure (d) is the error bar plot for apparent magnitude $m_b$ using the Pantheon 66 (40 bined plus 26 high red shift) SN Ia  $m_b$ data set.}
\label{Fig-4}
\end{figure}

\subsection{Plots of Deceleration, Jerks, Snap, Density and Pressure Parameters and Transitional Red Shifts on the basic of estimated parameter values:}
The deceleration parameter identifies the accelerating or decelerating phase of the model. In the fig. 5(a) it can be observed that there is a phase transition from the decelerating phase to the accelerating phase. The transitional red shifts, i.e., $ z_{tr} $, are approximately at  -0.742859, -0.78868, and  -0.61337 for each of the three data sets. So, the universe is at present in an accelerating phase, and before $ z_{tr} $,  it was decelerating.

The behavior of the jerk and snap parameters may also be analyzed from Figures 5(b) and 5(c). As per expectations, the jerk parameter is always positive, and the snap parameter shows transitional behavior like the deceleration parameter. The present values of the jerk parameters for the three data sets are different from  one. This shows that our model differs from the  $\Lambda$ CDM one. It will also be clear from the state finder diagnostic. The three curves in plots 5(a), 5(b), and 5(c) correspond to the three data sets used for estimating model parameters which are displayed in table-1.


\begin{figure}[H]
a.\includegraphics[width=6cm,height=6cm,angle=0]{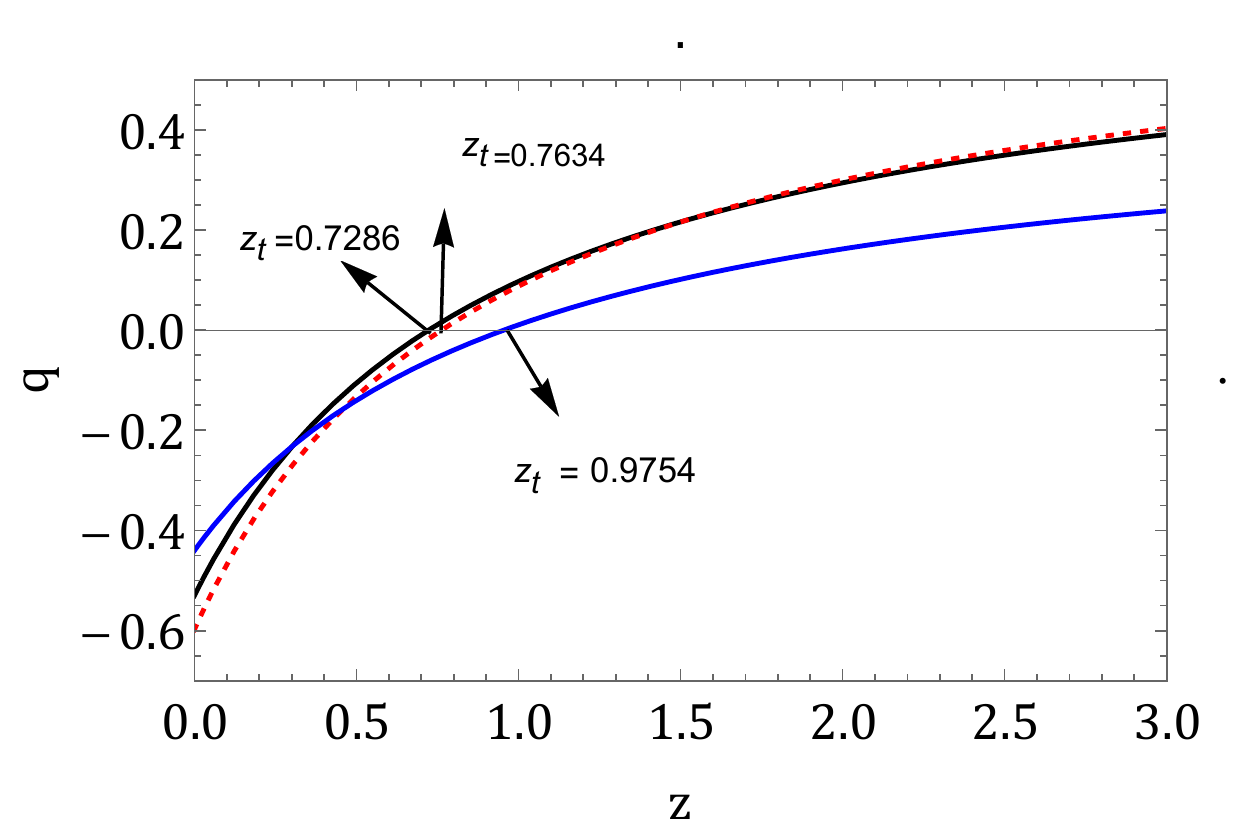}
	b.\includegraphics[width=6cm,height=6cm,angle=0]{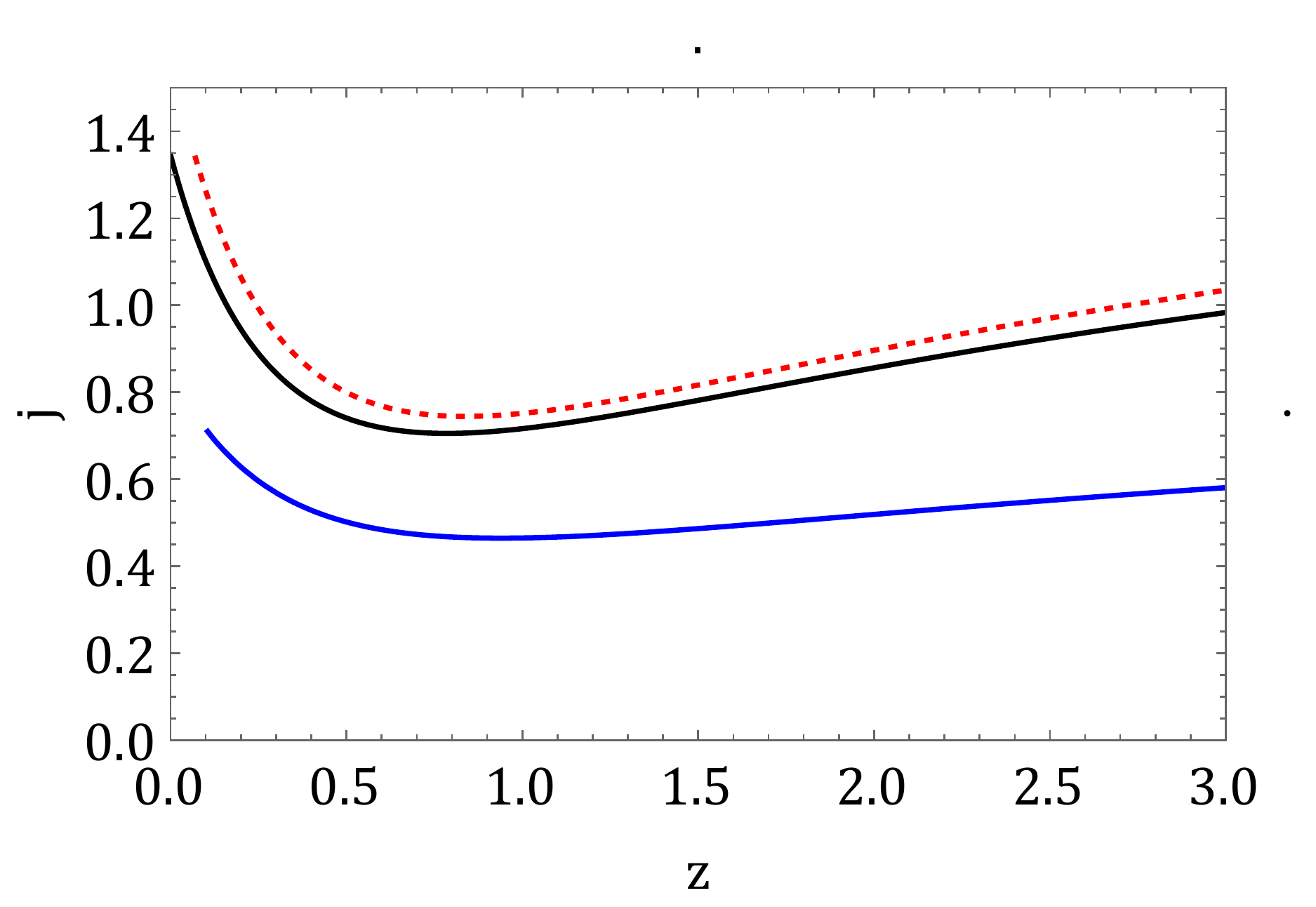}
	c.\includegraphics[width=6cm,height=6cm,angle=0]{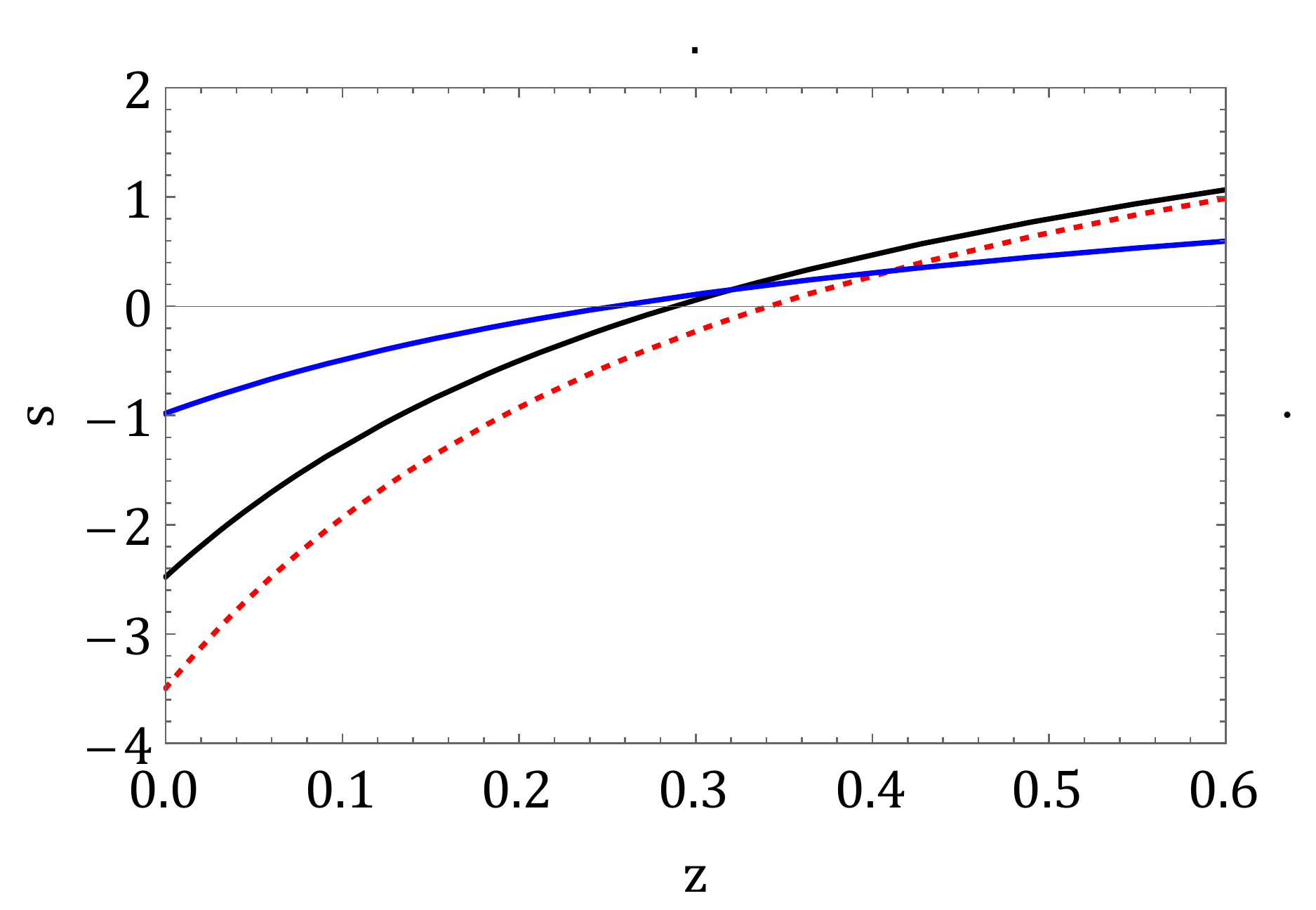}
\caption{ The plots of deceleration parameter $( q )$, jerk parameter $( j )$, snap parameter $( s )$. The three curves in plots $(a)$, $(b)$, and $(c)$ correspond  to the estimated values of model parameters for the three data sets displayed in table-1. }
\end{figure}
\subsection{Density and Pressure: }
From Eqs. (\ref{9}) and (\ref{10}) we get,
\begin{equation}\label{24}
	2 \dot{H} = -(8 \pi + 2 \lambda) ( p + \rho)
\end{equation}
Differentiating Eq. (\ref{12}) and using Eq. (\ref{24}), we get the energy conservation like equation in f(R,T) gravity as follows:
\begin{equation}\label{25}
	(8 \pi + 3 \lambda) \dot{\rho} - \lambda \dot{p}+ (8 \pi + 2 \lambda) 3 H (p + \rho) = 0
\end{equation}
Using $p=\omega \rho$, $\omega =\frac{ \omega_0}{1+z}$ and $\frac{ a_0}{a} = (1+z)$, we get the following differential equation  for $ \rho $ as a function of $z$.
\begin{equation}{\label{26}}
	\frac{\rho_z}{\rho} = \frac{3 (1+2 \eta) \bigg(1+\frac{3- \frac{k}{\eta}}{(1+z)} \bigg)  -  \eta \bigg( \frac { 3- \frac{k}{\eta}}{1+z}\bigg)}{(1+3 \eta) (1+z) - (3- \frac{k}{\eta} ) \eta}
\end{equation}
Solving this, we get the following expression for the energy density:
\begin{equation}\label{27}
	\rho (z) = \rho_0 \frac{ (1+\frac{(3 \eta+1)z}{1+k})^{\frac{1}{3 \eta +1}+\frac{3}{\eta }+7}}{(z+1)^{\frac{3}{\eta }+5}}.
\end{equation}

\begin{equation}\label{27a}
	p(z) = \omega_0\rho_0 \frac{ (1+\frac{(3 \eta+1)z}{1+k})^{\frac{1}{3 \eta +1}+\frac{3}{\eta }+7}}{(z+1)^{\frac{3}{\eta }+6}}.
\end{equation}
The behavior of the density parameter is  displayed in Figure 6(a). The variation of density $\rho$  over red shift $z$ indicates that in the past, the density was higher, but due to expansion, it decreased with time. The red shift is regarded as reciprocal of time. Figure 6(b) displays the behavior of pressure $p$ in the model. It is negative and anti-gravitating,  so produces acceleration in the universe. $\rho_0$ is computed as $\simeq 1.5 *10^{-29} gram/cm^3 $ where as the present value of pressure$p_0$ is computed as  $\simeq - 0.7 \rho_0$. The three curves in each figure correspond to the three data sets used for estimating model parameters which are displayed in Table -1. 

\begin{figure}[H]
\centering
(a){\includegraphics[width=6cm,height=6cm,angle=0]{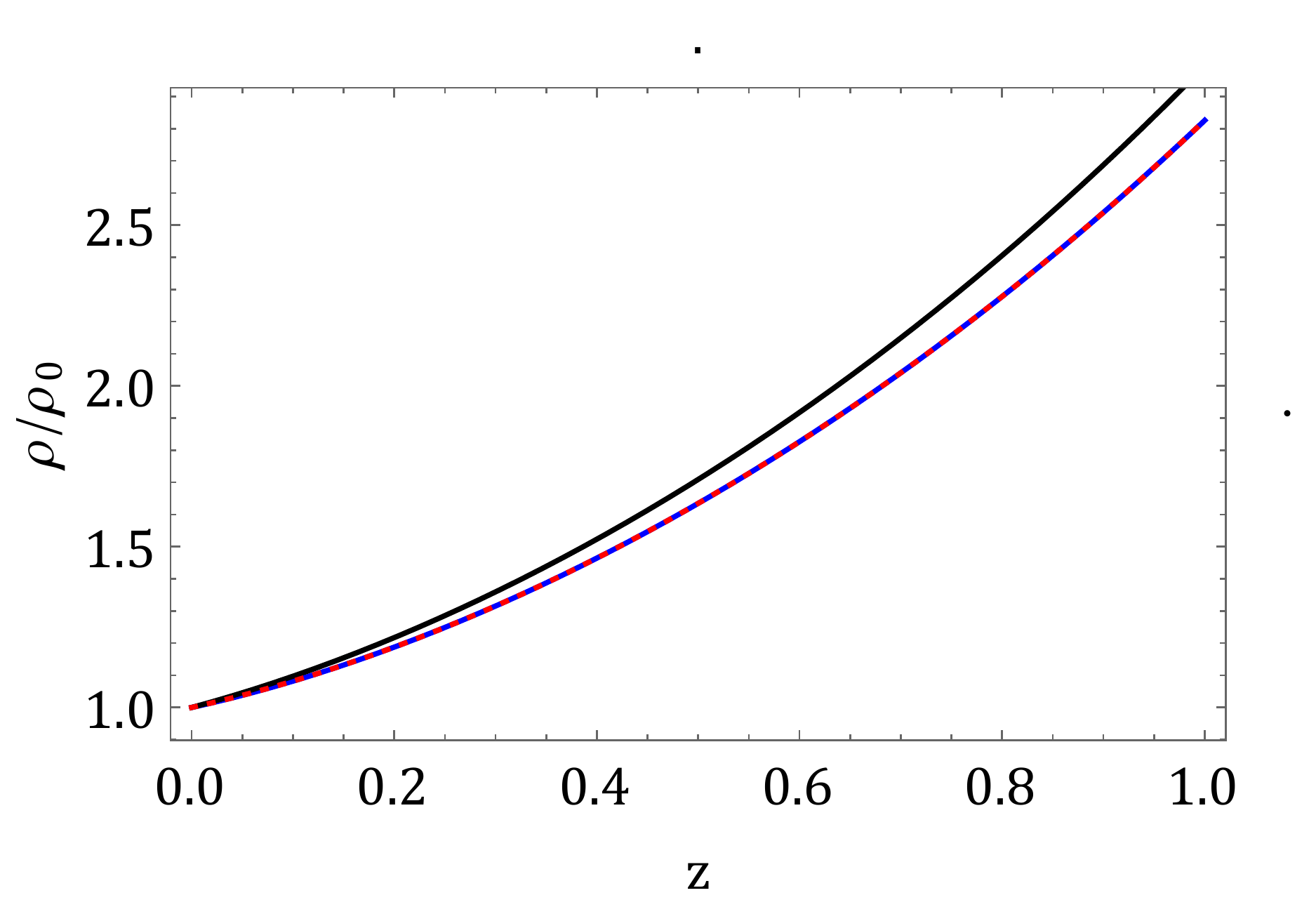} }
(b){\includegraphics[width=6cm,height=6cm,angle=0]{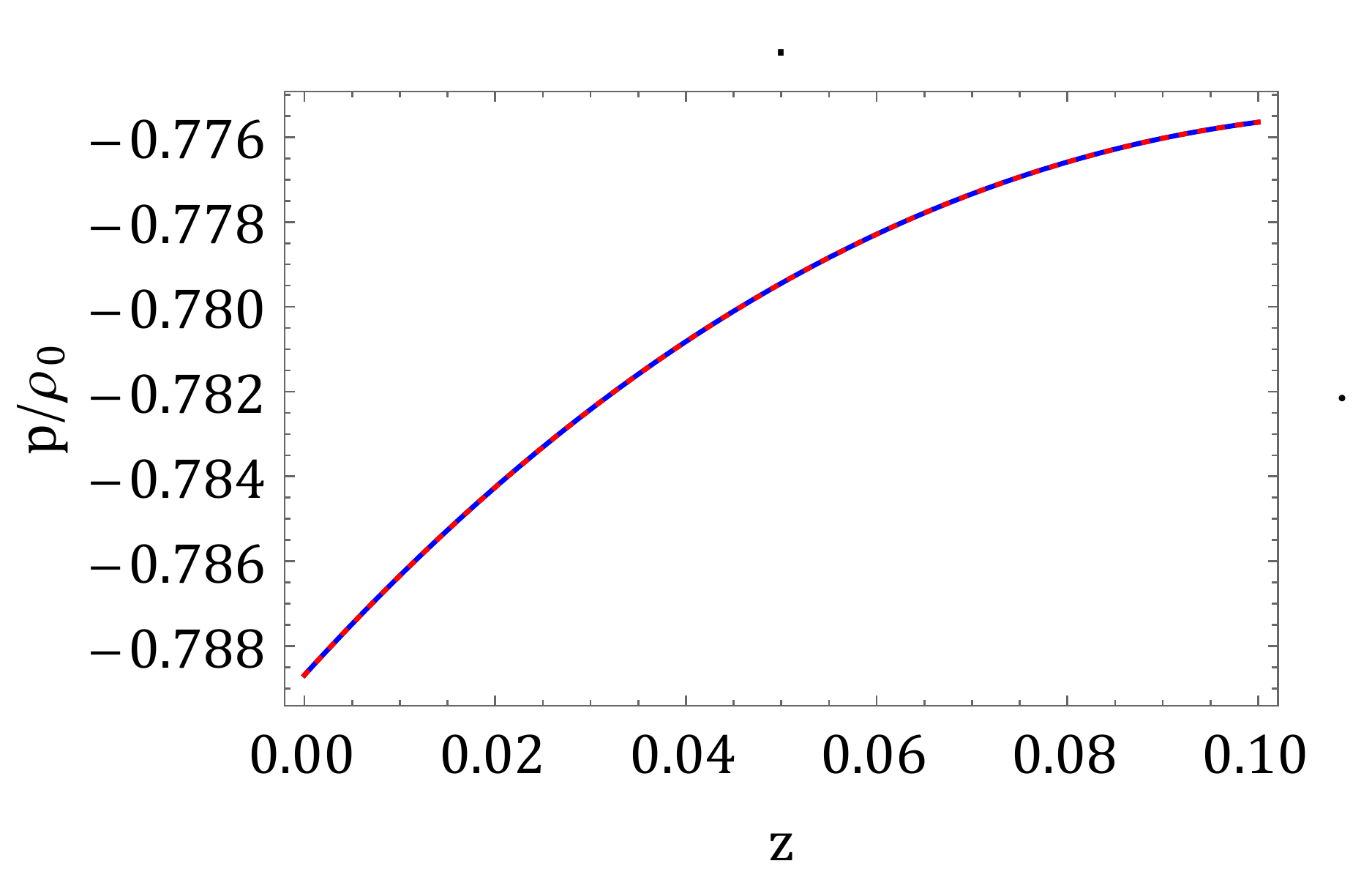}}
\caption{ The variation of density $\rho$ and pressure $p$ over red shift $z$. }
\end{figure}
\subsection{Time versus red shift Relation and Transitional times :}
\begin{figure}[H]
\centering
\includegraphics[width=6cm,height=6cm,angle=0]{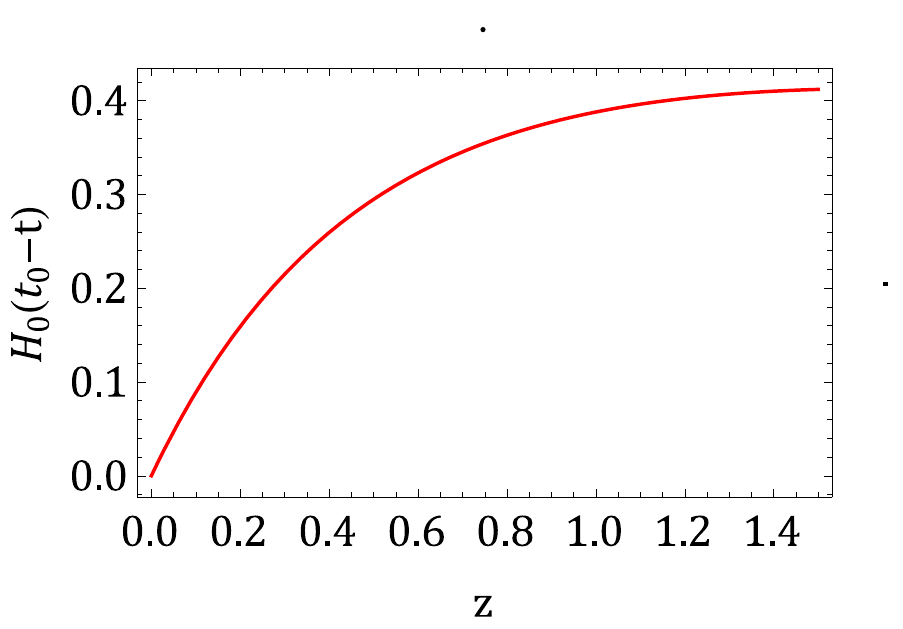}
\caption{The variation of red shift $z$ over time $t$.}
\label{Fig-7}
\end{figure}

The functional form of time over red shift is obtained from the following relation:
$$(t_0-t_1)=\intop_{t_1}^{t_0}dt=\intop_{a_1}^{a_0}\frac{da}{aH}=\intop_{0}^{z_1}\frac{dz}{(1+z)H(z)},$$
where we have used $\frac{a_0}{a}=1+z $ and $ \dot{z}=-(1+z)H.$ The values $t_0$ and $t_1$ are the present and some past time, respectively. We note that at present, $z=0$. From this, we can calculate the transitional time for which our universe entered into an accelerating phase. We have estimated that the transitional redshifts $z_t$ for the three data sets are $0.357056$, $0.346269$ and $0.377632$ respectively, so that  the corresponding   times $ (t_0-t_{z_t})$ are computed as $0.35705$ $H_0^{-1}$, $0.346269$ $H_0^{-1}$ and $0.377632$ $H_0^{-1}$yrs, respectively. Now  $H_0^{-1}= 9.8 * h_0^{-1} *10^{9}$ years, so the transitional time $t_{z_t}$  is calculated as $5.0061$, $5.52358$ and $5.03041$ billion years, respectively as of from now.  Figure (\ref{Fig-7}) displays the variation of red shift $z$ over time $t$.

\section{Statefinder diagnostic}

A geometrical diagnostic was used by Sahni {\it et al.} \cite{v} to introduce the pair of state finder parameters $(r, s)$ depending on scale factor "$a$". To distinguish a cosmological model from the standard $\Lambda$CDM model, one uses the $(r, s)$ analysis. The definition of the state finder pair is
$$s(\text{z})=\frac{r(z)-1}{3 \left(q(z)-\frac{1}{2}\right)}$$
We present two  plots of $s$ versus $r$, and $q$ versus $r$. Fig 8(a) shows that in all three estimations through the  $46$ OHD (Observational Hubble Data), SNIa $715$ DM($\mu$) data set and $66$ Pantheon AM $m_b$ data set, $s-r$ plots are meeting at $\Lambda$CDM point $(1,0)$ from both ends, i.e, from the Chaplygin gas model to the  $\Lambda$CDM and from quintessence to the $\Lambda$CDM stages.
The analysis of Fig 8(b) shows that our model is in quintessence at present, and its evolution passed through the $\Lambda$CDM and the Einstein-De Sitter stages in the past.

\begin{figure}[H]
	a.\includegraphics[width=9cm,height=8cm,angle=0]{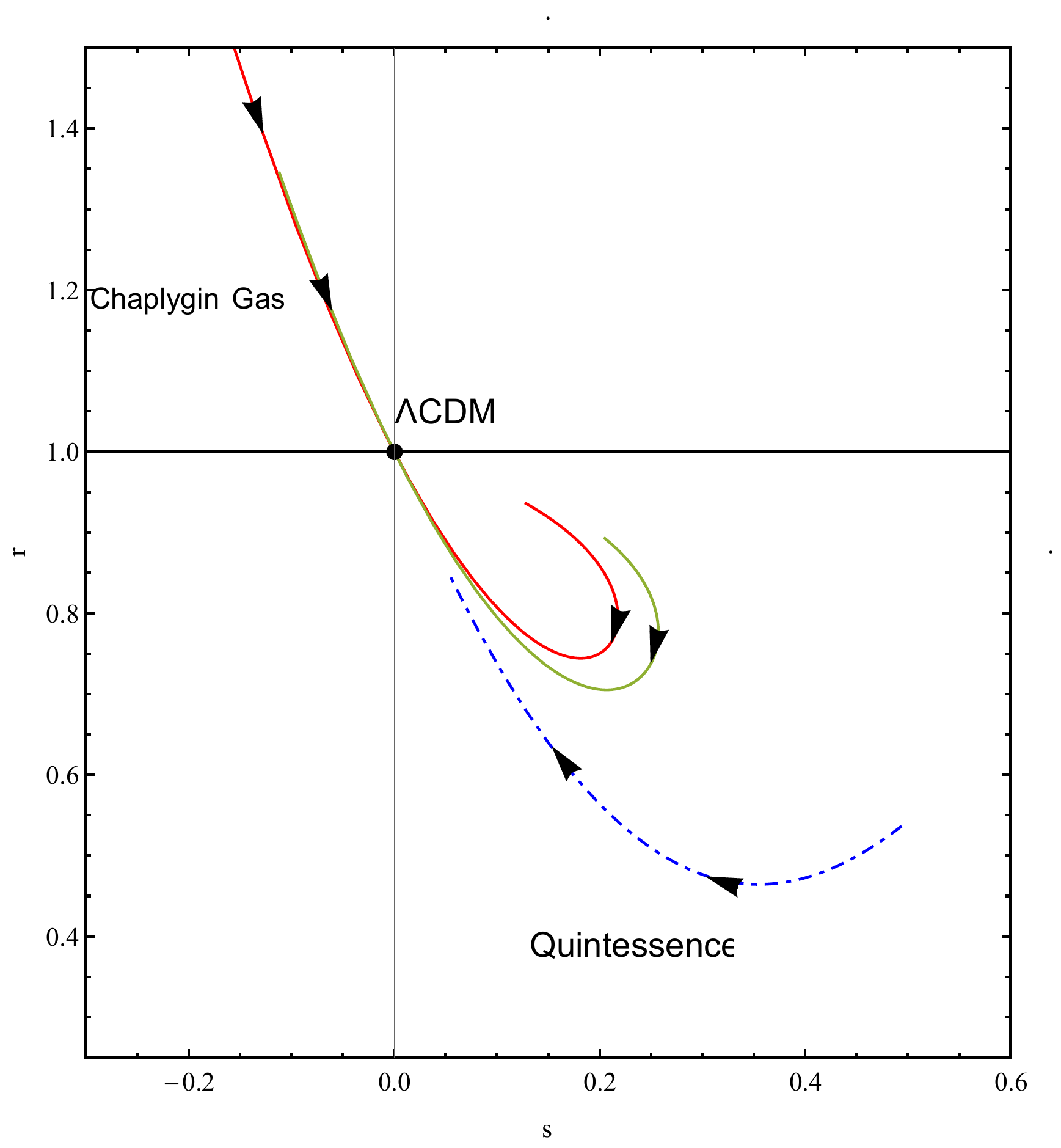}
	b.\includegraphics[width=9cm,height=8cm,angle=0]{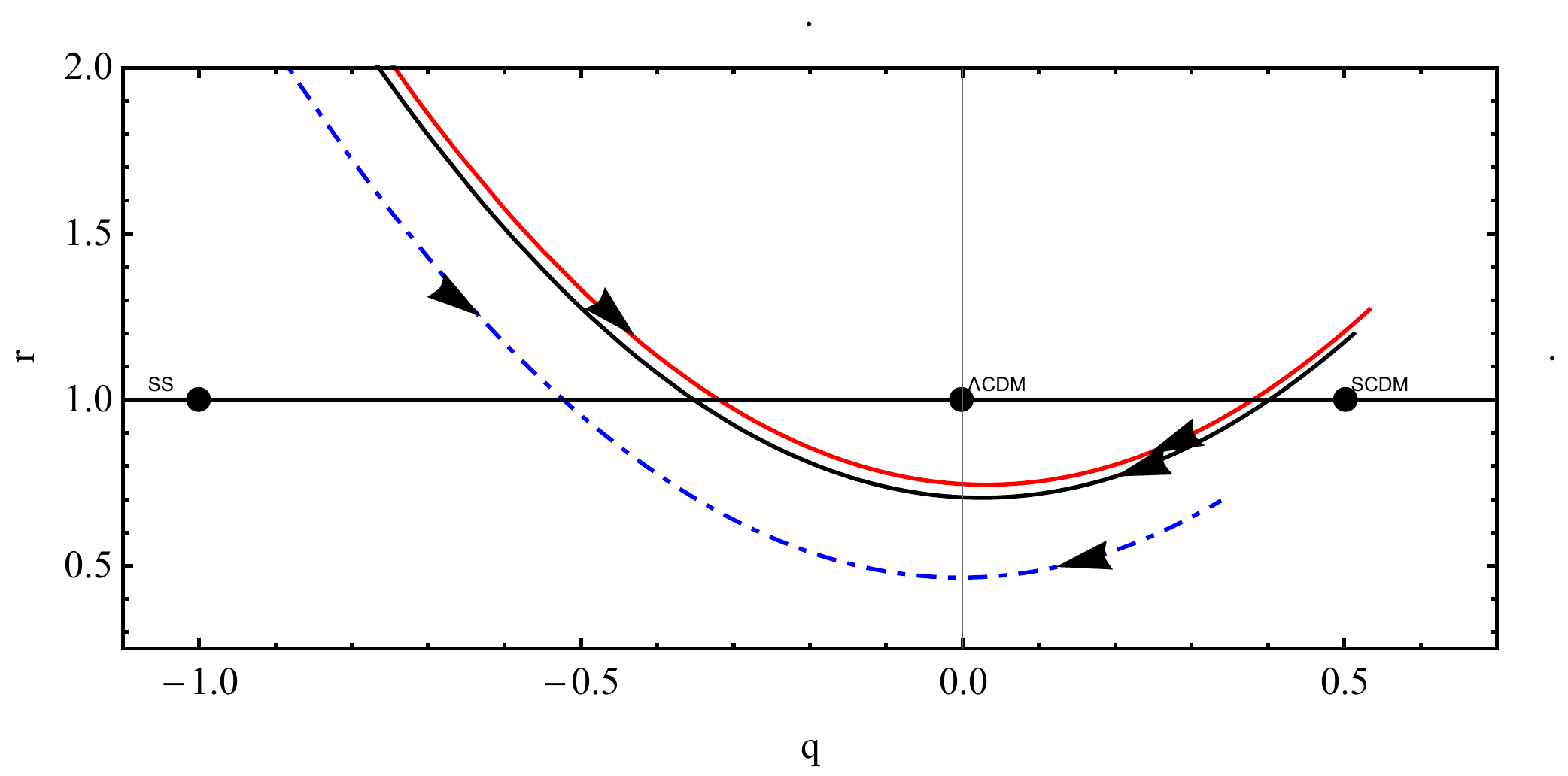}
\caption{ Dynamic behavior of our model universe in $(s,r)$ and $(q,r)$ planes.}
\end{figure}
\section{Conclusion:}
In this paper, we have modeled a universe in the framework of the FLRW metric using the field equations of $ f(R, T) $ gravity.  
\begin{enumerate}
    \item Efforts have been made to develop a cosmological model of the universe in f(R,T) gravity which is based on observational ground. For this, we have taken simplest functional   form $f(R, T) = R + 2 \lambda T$ where $\lambda$ is an arbitrary constant and  have assumed the ansatz for the equation of state parameter $\omega$ to solve the field equations with more than 2 unknowns.
    \item We have approximated the current values of model parameters i.e. Hubble$H_0$, Deceleration$q_0$,Equation of state$\omega_0$ and $\lambda$ in three ways with the help of three different Datasets:: $46$ OHD, SNeIa $715$ Distance modulus($\mu$), and $66$ Pantheon apparent magnitude $m_b$ high red shift data. The estimated calculated values lie near the present observed values of the Hubble and the Deceleration parameter.
    \item Then we have calculated the transitional red shifts and times for the data sets. The estimated value of density $\rho_0$ is $ \simeq 1.5 \rho_c $. The critical density is estimated as $ \rho_c\simeq 1.88~ h_0^2~10^{-29}~gm/cm^3 $ in the literature. The higher value of the present density is attributed to the presence of some additional energies in the universe apart from baryon energy. We have also examined the behavior  of the pressure in our model. It is negative and  is dominant over the density value  $\simeq - 0.7 \rho_0$.\\
    \item  We have carried out state finder analysis for our model which describes that 
     our model is a quintessence one at present, and its evolution passed through the $\Lambda$CDM and the Einstein-De Sitter stages in the past.
\end{enumerate}
We present the following table, which displays our estimated findings.
\begin{table}[H]
		\begin{tabular}{|c|c|c|c|c|c|c|c|c|c|}
		\hline
		S.No. & Data set type & $H_0$ & $q_0$ &$z_t$& $t_{z_t}$ & 
		$\lambda$ &$\omega_0$ & $\rho_0$ & $\chi^2$ \\
	
		\hline 
		1 & Hubble  & 67.786 &  -0.530244 & 0.7286&5.0061 byrs & -0.003387 & -0.742859 & 1.46794$\rho_c$  &21.6233\\
		 &46 data\\
		\hline
		
		  2 &715 SNeIa  & 69.5679 &  -0.597875 & 0.7634& 5.52358 byrs & -0.00380862 &-0.78868 &1.56901 $\rho_c$ &  1220.51\\
		  &$\mu$ data \\ 
		  \hline
		   3 & 66 Pantheon &--- &  -0.440048 &0.9754& 5.03041 byrs & 0.001 &-0.61337&0.922792 $\rho_c$ &95.40. \\
		   & ~$m_b$ data \\
		  \hline
		\end{tabular}
	\caption{Estimated parameter values}
        \label{table-1}
	\end{table}

\newpage
\begin{center}
{\bf{Table for Hubble Parameter $H$ observational values for different red shifts  }}
\end{center}
\begin{table}[H]
	\centering
	\begin{tabular}{|c|c|c|c|c|c|c|c|c|c|}
		\hline
		``S.No.& z & H(Obs) & $\sigma_i$ & References & S.No. & z & H(Obs) & $\sigma_i$ & References \\
		\hline 
		1 & 0     & 67.77 & 1.30 &  \cite{DES:2018rjw}     & 24 & 0.4783 & 80.9 & 9    &  \cite{Moresco::2016} \\
		\hline
		2 & 0.07 & 69    & 19.6  &  \cite{Zhang:2012mp}    & 25 & 0.48    & 97    & 60  & \cite{Stern:2010}  \\
		\hline
		3 & 0.09 & 69    & 12    &  \cite{Simon:2004tf}      & 26 & 0.51    & 90.4 & 1.9 & \cite{Chuang:2012qt}   \\
		\hline
		4 & 0.01 & 69    & 12    &  \cite{Stern:2010}      & 27 & 0.57    & 96.8 & 3.4 & \cite{Sarmah:2022hmf} \\
		\hline
		5 & 0.12 & 68.6 & 26.2 &  \cite{Zhang:2012mp}       & 28 & 0.593  & 104  & 13  &   \cite{Morsco::2012}  \\
		\hline
		6 & 0.17 & 83    & 8     &   \cite{Stern:2010}       &  29  &  0.60    & 87.9 & 6.1 &  \cite{Blake:2012pj}  \\
		\hline
		7 & 0.179 &  75 & 4 & \cite{Morsco::2012}   & 30 & 0.61 & 97.3 & 2.1 &  \cite{Chuang:2012qt}  \\
		\hline 
		8 & 0.1993 & 75 & 5 &  \cite{Morsco::2012}  & 31 & 0.68  & 92 & 8  &   \cite{Morsco::2012} \\
		\hline 
		9 & 0.2 & 72.9 & 29.6 &  \cite{Zhang:2012mp}  &  32 & 0.73 & 97.3 & 7 &  \cite{Blake:2012pj}  \\
		\hline 
		10 & 0.24 & 79.7 & 2.7 & \cite{Gaztanaga:2008xz}  & 33 & 0.781 & 105 & 12 &   \cite{Morsco::2012} \\
		\hline
		11 & 0.27 & 77 & 14 &  \cite{Stern:2010}  & 34  & 0.875 & 125 & 17 &   \cite{Morsco::2012}  \\
		\hline
		12 & 0.28 & 88.8 & 36.6 &  \cite{Zhang:2012mp}  & 35 & 0.88 & 90 & 40 & \cite{Stern:2010}   \\
		\hline 
		13 & 0.35 & 82.7 & 8.4 & \cite{BOSS:2016wmc}  & 36 & 0.9 & 117 & 23 &   \cite{Stern:2010}  \\
		\hline
		14 & 0.352 & 83 & 14 &   \cite{Morsco::2012}  & 37 & 1.037 & 154 & 20 & \cite{Gaztanaga:2008xz}   \\
		\hline 
		15 & 0.38 & 81.5 & 1.9 & \cite{Chuang:2012qt}  & 38 & 1.3 & 168 & 17 &   \cite{Stern:2010}  \\
		\hline 
		16 & 0.3802 & 83 & 13.5 &  \cite{BOSS:2016wmc}   & 39 & 1.363 & 160 & 33.6 &  \cite{Moresco:2015} \\
		\hline
		17 & 0.4 & 95 & 17 & \cite{Simon:2004tf}   & 40 & 1.43 & 177 & 18 &  \cite{Stern:2010}  \\
		\hline 
		18 & 0.4004 & 77 & 10.2 &  \cite{Moresco::2016} & 41 & 1.53 & 140 & 14 &  \cite{Stern:2010}  \\
		\hline 
		19 & 0.4247 & 87.1 & 11.2 & \cite{Moresco::2016} & 42 & 1.75 & 202 & 40 &  \cite{Moresco:2015}   \\
		\hline
		20 & 0.43 & 86.5 & 3.7 & \cite{Gaztanaga:2008xz}   & 43 & 1.965 & 186.5 & 50.4 & \cite{Gaztanaga:2008xz}  \\
		\hline 
		21 & 0.44 & 82.6 & 7.8 &  \cite{Blake:2012pj}   & 44 & 2.3   & 224 & 8 & \cite{Delubac::2013}   \\
		\hline
		22 & 0.497 & 92.8 & 12.9 & \cite{Moresco::2016}   & 45  & 2.34 & 222 & 7 & \cite{Delubac:2015}  \\
		\hline 
		23 & 0.47 & 89 & 49.6 & \cite{Ratsimbazafy:2017vga}  & 46 & 2.36 & 226 & 8 & \cite{Font-Ribera:2014}''  \\
		\hline  
	\end{tabular}
	\caption{The Hubble data}
	\label{Hubdata}
\end{table}


\section*{Acknowledgement}
The facilities provided by  IUCAA, Pune, India, during a visit when a part of this work was completed, are acknowledged and appreciated by the authors (A. Pradhan $\&$ G. K. Goswami). 

\section*{Conflicts of Interest:}
The authors declare no conflict of interest.


\begin{thebibliography}{99}
		\bibitem{a}   E. J. Copeland, M. Sami, and S. Tsujikawa,
	Dynamics of dark energy, Int. J. Mod. Phys. D {\bf 15} (2006)
	1753-1936.
	\bibitem{b}
	M. Li, X.-D. Li, S. Wang, and Y. Wang,
	Dark energy, Commun. Theor. Phys. {\bf 56} (2011) 525-604.
	\bibitem{c}
	O. Gron and S. Hervik, Einstein's general theory of relativity with modern applications in cosmology (Springer Publication, 2007)
	\bibitem{d}
	S. Weinberg,The cosmological constant problem, Rev. Mod. Phys. {\bf 61} (1989) 1.
	\bibitem{1}
	A.~G.~Riess, \textit{et al.} [Supernova Search Team],
	Observational evidence from supernovae for an accelerating universe and a cosmological constant, Astron. J. {\bf 116} (1998) 1009-1038.
	
		\bibitem{2}
	S.~Perlmutter, \textit{et al.} [Supernova Cosmology Project],
	Measurements of $\Omega$ and $\Lambda$ from 42 high redshift supernovae, Astrophys. J. {\bf 517} (1999) 565-586.
	
	\bibitem{3} S. Perlmutter, {\it et al.}, Discovery of a supernova explosion at half the age of the Universe, Nature {\bf 391} (1998) 51.
	
  
	\bibitem{4}
	J. L. Tonry, {\it et al.}, Cosmological results from high-z supernovae, Astrophys. J. {\bf 594} (2003) 1.
	
	\bibitem{5}
	A. Clocchiatti, {\it et al.}, Hubble Space Telescope and ground-based observations of type Ia Supernovae at redshift 0:5: cosmological implications, Astrophys. J. {\bf 642} (2006) 1.
	
	\bibitem{6}
	P. de Bernardis, {\it et al.}, A flat universe from high-resolution maps of the cosmic microwave background radiation, Nature {\bf 404} (2000) 955-959.
	
	\bibitem{7}
	S. Hanany, {\it et al.}, MAXIMA-1: a measurement of the cosmic microwave background anisotropy on angular scales of 10'$-$ 5, Astrophys. J. {\bf 545} (2000) L5$-$L9.
	
	\bibitem{8}
	D. N. Spergel, {\it et al.} [WMAP collaboration], First year wilkinson microwave anisotropy probe (WMAP) observations determination of cosmological parameters, Astrophys. J. Suppl. {\bf 148} (2003) 175.
	
	\bibitem{9}
	M. Tegmark, {\it et al.} [SDSS collaboration], Cosmological parameters from SDSS and WMAP, Phys. Rev. D {\bf 69} (2004) 103501.
	
	\bibitem{10}
	U. Seljak, {\it et al.}, Cosmological parameter analysis including SDSS Ly$\alpha$ forest and galaxy bias constraints on the primordial spectrum of fluctuations neutrino mass and dark energy, Phys. Rev. D {\bf 71} (2005) 103515.
	
	\bibitem{11}
	J. K. Adelman-McCarthy, {\it et al.}, The fourth data release of the sloan digital sky survey, Astrophys. J. Suppl. {\bf 162} (2006) 38.
	
	\bibitem{12}
	C. L. Bennett, {\it et al.}, First year wilkinson microwave anisotropy probe (WMAP) observations preliminary maps and basic results, The Astrophys. J. Suppl. {\bf 148} (2003) 1-43.
	
	\bibitem{13}
	S. W. Allen, {\it et al.}, Constraints on dark energy from chandra observations of the largest relaxed galaxy clusters, Mon. Not. R. Astron. Soc. {\bf 353}, (2004) 457.
	
	\bibitem{14}
	N. Suzuki, {\it et al.}, The Hubble space telescope cluster supernova survey V improving the darkenergy constraints above z $>$ 1 and building an early-type-hosted supernova sample, Astrophys. J. {\bf 746} (2012) 85-115.
	
	\bibitem{15}
	T. Delubac, {\it et al.} [BOSS Collaboration], 2015. Baryon acoustic oscillations in the Ly$\alpha$ forest of BOSS DR11 quasars, Astron. Astrophys. {\bf 574} (2015) A59.
	
	\bibitem{16}
	C. Blake, {\it et al.} [The Wiggle Z Dark Energy Survey], 2012. The Wiggle Z dark energy survey joint measurements of the expansion and growth history at $z < 1$ , Mon. Not. R. Astron. Soc. {\bf 425} (2012) 405-414.
	
	\bibitem{17}
	P. A. R. Ade, {\it et al.} [Planck Collaboration], 2016. Planck 2015 results XIV dark energy and modied gravity, Astron. Astrophys. {\bf 594} (2016) A14.
	
		\bibitem{18}
	E.~Komatsu, \textit{et al.} [WMAP], Five-year wilkinson microwave anisotropy probe (WMAP) observations: Cosmological interpretation,
	Astrophys. J. Suppl. \textbf{180} (2009) 330-376.
	
	\bibitem{19}
	E.~Komatsu, \textit{et al.}, Seven-Year Wilkinson Microwave Anisotropy Probe (WMAP) Observations: Cosmological Interpretation, Astrophys. J. Suppl. \textbf{192} (2011) 18.
	
	\bibitem{20}
	N.~Aghanim \textit{et al.}, Planck 2018 results. VI. Cosmological parameters,
	Astron. Astrophys. \textbf{641} (2020) A6. [erratum: Astron. Astrophys. \textbf{652} (2021) C4.
	
	\bibitem{21}
	S. Alam {\it et al.} (BOSS Collaboration), The clustering of galaxies in the completed
	SDSS-III Baryon Oscillation Spectroscopic Survey: cosmological analysis of the
	DR12 galaxy sample, Mon. Not. Roy. Astron. Soc. {\bf 470} (2017) 2617-2652.

	
	\bibitem{22} M. M. Ivanov, Cosmological constraints from the power spectrum of eBOSS
	emission line galaxies, Phys. Rev. D {\bf 104} (2021) 103514.
	
		\bibitem{23} O. H. E. Philcox, M. M. Ivanov, M. Simonovic, and M. Zaldarriaga, Combining full-shape and BAO analyses of galaxy power spectra: A 1.6\%
	CMB-independent constraint on H0, JCAP {\bf 2020(05)} (2020) 032.
	
		\bibitem{24} O. H. E. Philcox, B. D. Sherwin, G. S. Farren, and E. J. Baxter, Determining the Hubble constant without the sound horizon: Measurements from galaxy surveys, Phys. Rev. D {\bf 103} (2021) 023538.
	
		\bibitem{25} T. Colas, G. D'amico, L. Senatore, P. Zhang, and F. Beutler, Efficient cosmological analysis of the SDSS/BOSS data from the effective field theory of large-scale structure, JCAP {\bf 2020(06)} (2020) 001.
	\bibitem{e}
	P.~J.~E.~Peebles and B.~Ratra, The Cosmological constant and dark energy,
	Rev. Mod. Phys. \textbf{75} (2003) 559-606.
	
	 \bibitem{f}
	P. Steinhardt, L. Wang and I. Zlatev, Cosmological tracking solutions, Phys. Rev. D {\bf 59} (1999) 123504.
	\bibitem{g}
	V. B. Johri, Genesis of cosmological tracker fields, Phys. Rev. D {\bf 63} (2001) 103504.
		\bibitem{g1}
H. Motohashi, A. A. Starobinsky, and J. Yokoyama, f(R) gravity and its cosmological implications, Int. J. Mod. Phys. D {\bf 20} (2011) 1347-1355.
	
\bibitem{h}
	J.~Frieman, M.~Turner, and D.~Huterer,
	Dark energy and the accelerating universe,
	Ann. Rev. Astron. Astrophys. \textbf{46} (2008) 385-432 (2008).

\bibitem{i}	G. F. R. Ellis, U. Kirchner, and W. R. Stoeger,
	Multiverses and physical cosmology, Mon. Not. R. Astron. Soc. {\bf 347} (2004) 921-936.

    \bibitem{j}
    S.~Nojiri and S.~D.~Odintsov,
    Introduction to modified gravity and gravitational alternative for dark energy,
    eConf \textbf{C0602061} (2006) 06, [arXiv:hep-th/0601213].
    
    \bibitem{k}
    T.~P.~Sotiriou and V.~Faraoni,
    f(R) theories of gravity, Rev. Mod. Phys. \textbf{82} (2010) 451-497.

    \bibitem{l}
    F.~S.~N.~Lobo, The Dark side of gravity: Modified theories of gravity,
    [arXiv:0807.1640 [gr-qc]].
    
    \bibitem{m}
    S.~Capozziello and M.~Francaviglia,
    Extended theories of gravity and their cosmological and astrophysical applications,
    Gen. Rel. Grav. \textbf{40} (2008) 357-420.
    
    \bibitem{n}
    S.~M.~Carroll, V.~Duvvuri, M.~Trodden and M.~S.~Turner,
    Is cosmic speed-up due to new gravitational physics?,
    Phys. Rev. D \textbf{70} (2004) 043528.

    \bibitem{o}
    S. Nojiri, S. D. Odintsov, and O. G. Gorbunova,
    Dark energy problem: from phantom theory to modified Gauss-Bonnet gravity,
    Journal of Physics A: Mathematical General, \textbf{39} (2006) 6627.
    
    \bibitem{p} A. Chudaykin, K. Dolgikh, and M. M. Ivanov, Constraints on the curvature of the universe and dynamical dark energy from the full-shape and BAO data, Phys.
	Rev. D {\bf 103} (2021) 023507.
	

	
	

	
	\bibitem{s}
	L. D. Landau and E. M. Lifshitz, The Classical Theory of Fields, Butterworth-Heinemann, Oxford (1998).
	
	\bibitem{t}
	T.~Harko, F.~S.~N.~Lobo, S.~Nojiri, and S.~D.~Odintsov,
	$f(R,T)$ gravity, Phys. Rev. D \textbf{84} (2011) 024020.

	\bibitem {t1}
K. S. Adhav, LRS Bianchi type-I cosmological model in $f(R,T)$ theory of gravity, Astrophys. Space Sci. {\bf 339} (2012) 365.
\bibitem {t2}
G. C. Samanta, Universe filled with dark energy (DE) from a wet dark fluid (WDF) in $f(R,T)$ gravity, Int. J. Theor. Phys.
{\bf 52} (2013) 2303.
\bibitem {t3}
G. C. Samanta and S. N. Dhal, Higher dimensional cosmological models filled with perfect fluid in $f(R,T)$ theory of gravity,
Int. J. Theor. Phys. {\bf 52} (2013) 1334.
\bibitem {t4}
M. F. Shamir, Bianchi type I cosmology in $f(R,T)$ gravity, J. Exp. Theor. Phys. {\bf 119} (2014) 242.
\bibitem {t5}
R. Chaubey and A. K. Shukla, A new class of Bianchi cosmological models in $f(R,T)$ gravity, Astrophys. Space Sci. {\bf 343} (2013) 415.

\bibitem {t6}
S. B. Fisher and E.D. Carlson, Reexamining $f(R,T)$ gravity, Phys. Rev. D {\bf 100} (2019) 064059.
\bibitem {t7}
T. Harko and P. H. R. S. Moraes, Comment on Reexamining $f(R,T)$ gravity, Phys. Rev. D {\bf 101} (2020) 108501.
\bibitem {t8}
G. A. Carvalho, R. V. Lobato, P. H. R. S. Moraes, J. D. V. Arbail, R. M. Marinho, J. E. Otoniel, and M. Malheiro, Stellar equilibrium configurations of
white dwarfs in the $f(R, T)$ gravity, Eur. Phys. J. C {\bf 99} (2017) 871.
\bibitem {t9}
T. M. Ordines and E. Carlson, Limits on $f(R,T)$ gravity from Earth’s atmosphere, Phys. Rev. D {\bf 99} (2019) 104052.
\bibitem {t10}
A. K. Yadav, P. K. Sahoo, and V. Bhardwaj, Bulk viscus Bianchi-I embedded cosmological model in $f(R, T) = f_{1}(R) + f_{2}(R)f_{3}(T)$ gravity,
Mod. Phys. Lett. A {\bf 34} (2019) 1950145.
\bibitem {t11}
L. K. Sharma, B. K. Singh, and A. K. Yadav, Viability of Bianchi type $V$ universe in  $f(R, T) = f_{1}(R) + f_{2}(R)f_{3}(T)$ gravity,
Int. J. Geom. Methods Mod. Phys. {\bf 17} (2020) 2050111.
\bibitem {t12}
V. K. Bhardwaj, M. K. Rana, and A. K. Yadav, Bulk viscous Bianchi-V cosmological model within the formalism of  $f(R, T) = f_{1}(R) + f_{2}(R)f_{3}(T)$
gravity, Astrophys. Space Sci {\bf 364} (2019) 136.
\bibitem {t13}
L. K. Sharma, A. K. Yadav, P. K. Sahoo, and B. K. Singh, Non-minimal matter-geometry coupling in Bianchi I space-time, Res. Phys. {\bf 10} (2018) 738.

\bibitem {t14}
V. K. Bhardwaj and A. Pradhan, Evaluation of cosmological models in f(R, T) gravity in different dark energy scenario, New Astronomy {\bf91} (2022) 101675.
\bibitem {t15}
T. Tangphati, S. Hansraj, A. Banerjee, and A. Pradhan, Quark stars gravity with an interacting quark equation of state, Phys. Dark Univ. {\bf 35} (2022) 100990.
\bibitem {t16}
J. M. Z. Pretel, T. Tangphati, A. Banerjee, and A. Pradhan, Charged quark stars in f(R, T) gravity, Chin. Phys. C {\bf 46} (2022) 115103.

\bibitem{u}
	Y.~G.~Gong and Y.~Z.~Zhang, Probing the curvature and dark energy,
	Phys. Rev. D \textbf{72} (2005) 043518.
\bibitem{v}
K. Bamba, S. Capozziello, S. Nojiri and S. D. Odintsov,
Dark energy cosmology: the equivalent description via different theoretical models and cosmography tests
Astrophys.\ Space Sci. \ {\bf 342} (2012), 155 
\bibitem{w}
S.~Nojiri and S.~D.~Odintsov,
Unified cosmic history in modified gravity: from F(R) theory to Lorentz non-invariant models,
Phys. Rept. \textbf{505} (2011) 59-144


\bibitem{x}
S.~Nojiri, S.~D.~Odintsov and V.~K.~Oikonomou,
Modified Gravity Theories on a Nutshell: Inflation, Bounce and Late-time Evolution,
Phys. Rept. \textbf{692} (2017), 1-104

\bibitem{DES:2018rjw}
E.~Macaulay, \textit{et al.} [DES], First cosmological results using Type Ia supernovae from the dark energy survey: Measurement of the Hubble constant,
Mon. Not. Roy. Astron. Soc. \textbf{486} (2019) 2184-2196.

\bibitem{Zhang:2012mp}
C.~Zhang, H.~Zhang, S.~Yuan, T.~J.~Zhang and Y.~C.~Sun,
four new observational $H(z)$ data from luminous red galaxies in the sloan digital sky Survey data release seven, Res. Astron. Astrophys. \textbf{14} (2014) 1221-1233.

\bibitem{Stern:2010}
D.~Stern, R.~Jimenez, L.~Verde, M.~Kamionkowski, SA.~Stanford,
Cosmic chronometers: constraining the equation of state of dark energy. I: H (z) measurements, JCAP \textbf{2010(02)} (2010) 008.

\bibitem{Gaztanaga:2008xz}
E.~Gaztanaga, A.~Cabre and L.~Hui,
Clustering of luminous red galaxies IV: Baryon acoustic peak in the line-of-sight direction and a direct measurement of H(z),
Mon. Not. Roy. Astron. Soc. \textbf{399} (2009) 1663-1680.

\bibitem{Chuang:2012qt}
C.~H.~Chuang and Y.~Wang,
Modeling the anisotropic two-point galaxy correlation Function on Small Scales and Improved Measurements of $H(z)$, $D_A(z)$, and $\beta(z)$ from the Sloan Digital Sky Survey DR7 Luminous Red Galaxies,
Mon. Not. Roy. Astron. Soc. \textbf{435} (2013) 255-262.



\bibitem{BOSS:2016wmc}
S.~Alam \textit{et al.} [BOSS],
The clustering of galaxies in the completed SDSS-III baryon oscillation spectroscopic survey: cosmological analysis of the DR12 galaxy sample,
Mon. Not. Roy. Astron. Soc. \textbf{470} (2017) 2617-2652.

\bibitem{Blake:2012pj}
C.~Blake, S.~Brough, M.~Colless, C.~Contreras, W.~Couch, S.~Croom, D.~Croton, T.~Davis, M.~J.~Drinkwater and K.~Forster, \textit{et al.}
The WiggleZ dark energy survey: Joint measurements of the expansion and growth history at z \ensuremath{<} 1, Mon. Not. Roy. Astron. Soc. \textbf{425} (2012) 405-414.


\bibitem{Ratsimbazafy:2017vga}
A.~L.~Ratsimbazafy, S.~I.~Loubser, S.~M.~Crawford, C.~M.~Cress, B.~A.~Bassett, R.~C.~Nichol and P.~V\"ais\"anen,
Age-dating luminous red galaxies observed with the Southern African large telescope,
Mon. Not. Roy. Astron. Soc. \textbf{467} (2017) 3239-3254.

\bibitem{Moresco:2015}
M. Moresco, Raising the bar: new constraints on the Hubble parameter with cosmic chronometers at $z \sim 2$, Mont. Not. Royal Astron. Soci. Lett. \textbf{450} (2015) L16-L20.

\bibitem{Simon:2004tf}
J.~Simon, L.~Verde, and R.~Jimenez,
Constraints on the redshift dependence of the dark energy potential,
Phys. Rev. D \textbf{71} (2005) 123001.

\bibitem{Morsco::2012} 
M. Moresco, {\it et al.}
 Improved constraints on the expansion rate of the universe up to $z \sim 1.1$ from the spectroscopic evolution of cosmic chronometers, JCAP \textbf{2012(08)} (2012) 006.

\bibitem{Moresco::2016}
M. Moresco, {\it et al.}
A 6 measurement of the Hubble parameter at $z \sim 0.45$: direct evidence of the epoch of cosmic re-acceleration,
JCAP \textbf{2016(05)} (2016) 014.

\bibitem{Sarmah:2022hmf}
P.~Sarmah and U.~D.~Goswami, Bianchi Type I model of universe with customized scale factors, arXiv:2203.00385 [gr-qc] (2022).

\bibitem{Delubac::2013}
T. Delubac, J. Rich, S. Bailey, A. Font-Ribera, {\it et al.},
Baryon acoustic oscillations in the Ly$\alpha$ forest of BOSS quasars,
Astron. Astrophys. \textbf{552} (2013) A96.

\bibitem{Delubac:2015}
T. Delubac, J. E. Bautista, J. Rich, D. Kirkby, {\it et al.}
Baryon acoustic oscillations in the Ly$\alpha$ forest of BOSS DR11 quasars,
Astron. Astrophys. \textbf{574} (2015) A59.

\bibitem{Font-Ribera:2014}  
A. Font-Ribera, {\it et al.},
Quasar-Lyman $\alpha$ forest cross-correlation from BOSS DR11: Baryon acoustic oscillations, JCAP \textbf{2014(05)} (2014) 027.

\bibitem{v} V. Sahni, T. D. Saini, A. A. Starobinsky and U. Alam, Statefinder- A new geometrical diagnostic of dark energy, JETP Lett. {\bf 77} (2003) 201.
\end{thebibliography}
\end{document}